\documentclass[apj]{emulateapj}
\usepackage{apjfonts}

\begin{document}

\newcommand{\boot}{Bo\"otes}
\newcommand{\cetus}{Cetus}
\newcommand{\chandra}{{\it Chandra}}
\newcommand{\galex}{{\it GALEX}}
\newcommand{\spitzer}{{\it Spitzer}}
\newcommand{\irac}{{IRAC}}
\newcommand{\mips}{{MIPS}}
\newcommand{\kms}{~km~s$^{-1}$}
\newcommand{\logh}{$+5\log h_{70}$}
\newcommand{\ho}{~}
\newcommand{\bootarea}{4.7}
\newcommand{\netbootarea}{4.55}
\newcommand{\totarea}{7.1}
\newcommand{\cetarea}{2.4}
\newcommand{\msun}{M$_\odot$~}
\newcommand{\degtwo}{deg$^2$}
\title{The FLAMINGOS Extragalactic Survey}

\author{
Richard J.~Elston\altaffilmark{1,2}, 
Anthony H.~Gonzalez\altaffilmark{1,3}, 
Eric McKenzie\altaffilmark{1}, 
Mark Brodwin\altaffilmark{4},
Michael J.~I.~Brown\altaffilmark{5,6},
Gustavo~Cardona\altaffilmark{1},
Arjun~Dey\altaffilmark{5},
Mark~Dickinson\altaffilmark{5},
Peter R.~Eisenhardt\altaffilmark{4},
Buell T.~Jannuzi\altaffilmark{5},
Yen-Ting Lin\altaffilmark{7},
Joseph J.~Mohr\altaffilmark{7,8},
S. Nicholas Raines\altaffilmark{1},
S.~A.~Stanford\altaffilmark{9,10},
Daniel Stern\altaffilmark{4}
}

\altaffiltext{1}{Department of Astronomy, University of Florida,
  Gainesville, FL 32611}
\altaffiltext{2}{Deceased}
\altaffiltext{3}{NSF Astronomy and Astrophysics Postdoctoral Fellow, anthony@astro.ufl.edu}
\altaffiltext{4}{Jet Propulsion Laboratory, California Institute of Technology, 
4800 Oak Grove Drive, Pasadena, CA 91109}
\altaffiltext{5}{National Optical Astronomy Observatory, Tucson, AZ, 85726-6732}
\altaffiltext{6}{Department of Astrophysical Sciences, Princeton University, Peyton Hall,
Princeton, NJ 08544-1001}
\altaffiltext{7}{Department of Astronomy, University of Illinois, Urbana, IL 61801}
\altaffiltext{8}{Department of Physics, University of Illinois, Urbana, IL 61801}
\altaffiltext{9}{Physics Department, University of California, Davis, CA 95616}
\altaffiltext{10}{Institute of Geophysics and Planetary Physics, Lawrence Livermore National 
Laboratory, Livermore, CA 94551}

\begin{abstract}

Using the Florida Multi-object Imaging Near-IR grism Observational Spectrometer (FLAMINGOS),
we have conducted the FLAMINGOS Extragalactic Survey (FLAMEX), a deep imaging survey covering
\totarea~$\deg^2$ within the 18.6 $\deg^2$ NOAO Deep Wide-Field Survey (NDWFS)
regions. FLAMEX is the first deep, wide-area near-infrared survey to image in
both the $J$ and $K_s$ filters, and is larger than any previous NIR surveys of
comparable depth. The intent of FLAMEX is to facilitate the study of galaxy and
galaxy cluster evolution at $1<z<2$ by providing rest-frame optical photometry
for the massive galaxy population at this epoch. This effort is designed to
yield a public data set that complements and augments the suite of existing
surveys in the NDWFS fields. We present an overview of FLAMEX and initial
results based upon $\sim150,000$ $K_s-$selected sources in the \boot~field.
We describe the observations and reductions, quantify the data quality, and
verify that the number counts are consistent with results from previous
surveys. Finally, we comment upon the utility of this sample for detailed
study of the ERO population, and present one of the first spectroscopically
confirmed $z>1$ galaxy clusters detected using the joint FLAMEX, NDWFS, and
\spitzer~\irac~Shallow Survey data sets.

\end{abstract}

\keywords{surveys --- infrared: galaxies --- galaxies: evolution --- galaxies: clusters }

\section{Introduction}

Two fundamental goals of extragalactic astronomy are to understand the star
formation and mass assembly histories of the universe.  It has become
increasingly apparent in the last few years that the epoch $z=1-2$ is critical
for the evolution of both quantities.  Current observations indicate that the
bulk of stellar mass assembly in galaxies and $\sim50$\% of the total star
formation occurs in this interval
\citep{Fontana2003,Dickinson2003,Rudnick2003}.  It is also during this epoch
that we expect to witness the formation of the first massive galaxy clusters
\citep[for example, see][]{Haiman2001}.  Observations at this epoch thus
promise to provide a portrait of the universe during its most vibrant
evolutionary period.  Yet, in constrast to the wealth of data at low redshifts
\citep[$z<1$,][]{York2000,Colless2001,Kochanek2004} and high redshifts
\citep[$z=2.5-4$,][]{Steidel1998,Steidel1999, Adelberger2003}, this redshift
regime remains less well explored due largely to the observational challenges
historically associated with the ``redshift desert'' at $z=1.4-2.5$ \citep[for
an overview, see][]{Steidel2004}.  Recent spectroscopic work by a number of
groups \citep[e.g.,][]{Abraham2004, Coil2004,
Steidel2004, Adelberger2005a,Adelberger2005b,Mignoli2005} has begun to open this redshift
window, but we currently lack large, homogenous samples comparable to those
that exist at other redshifts. Moreover, these existing programs are
restricted to relatively small areas and are thus unable to probe the fraction
of the massive galaxy population that resides in the overdense group and
cluster environments.  Infrared imaging surveys are a key first step towards
generation of such samples. Infrared observations (both near-infrared and with
\irac on the \spitzer~{\it Space Telescope}) have the advantage that the
combined evolutionary and $k-$corrections are mild at this epoch,
which means that uniform, roughly stellar mass-limited samples can be selected
over a range of redshifts.

Using the Florida Multi-object Imaging Near-IR grism Observational Spectrometer (FLAMINGOS),
we have conducted the FLAMINGOS Extragalactic Survey (FLAMEX). FLAMEX is
a wide-area, near-infrared (NIR) imaging survey that is motivated by
two basic considerations.
First, FLAMEX is designed to probe the evolution and clustering of the massive
galaxy population at $z=1-2$ and to identify a sample of galaxy clusters at
this epoch that can be used to study the assembly of cluster galaxies. The
NDWFS fields are the targets of one of the most extensive panchromatic
investigations in all astronomy, with space- and ground-based imaging and
spectroscopic programs spanning radio to X-ray wavelengths
\citep{Jannuzi1999,Rhoads2000,deVries2002,Hoopes2003,Lonsdale2003,Eisenhardt2004,Kochanek2004,Pierre2004,Houck2005,Murray2005}. The
northern (\boot) field in fact is the only wide-area survey region presently
mapped by \chandra, \galex, and \spitzer~(with \irac~and \mips).

\begin{figure}
\plotone{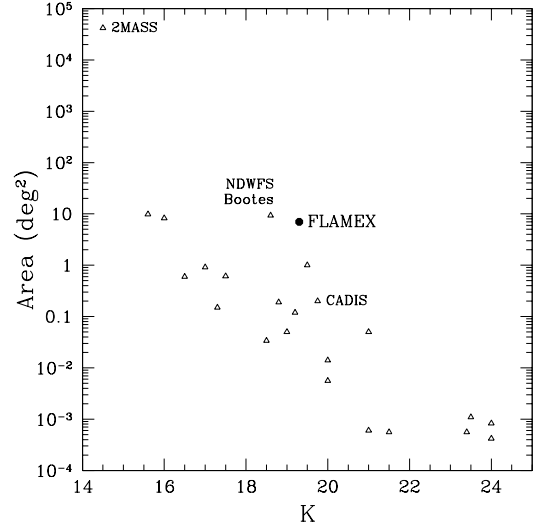}
\caption{Comparison of FLAMEX with previous $K-$band surveys.  The points show
the published limiting magnitudes and areas for all the surveys listed in
Table \ref{tab:surveys}, except for the $H-$band LCIRS.  Among the wide-area
surveys shown, only FLAMEX, 2MASS, and MUNICS provide complete coverage in both $J$ and $K_s$ data.  We note
that no correction has been applied to correct for the different filters ($K$,
$K^\prime$,$K_s$) used by the different surveys, and that different authors
quote different significance limiting magnitudes, so the data points provide
only an approximate comparison of survey depths.  \label{fig:surveycomparea}}
\end{figure}

FLAMEX provides a combination of area and depth in both $J$ and $K_s$ that is
unique among current NIR surveys (see Table \ref{tab:surveys} and Figure
\ref{fig:surveycomparea}).  Coupling the FLAMEX data with the optical (and
when available \irac) photometry enables derivation of robust photometric
redshifts (Brodwin et al., in prep., and Brown et al., in prep.), providing an
efficient means of isolating large samples of massive galaxies at
$z=1-2$. This combination of mass selection and photometric redshifts over a
wide area constitutes a powerful tool not only for galaxy cluster detection
and galaxy evolution studies, but also for selection of subsamples of rare
objects that can be studied in greater detail.

Second, the addition of FLAMEX to the NDWFS repertoire provides the requisite
deep NIR photometry that will be required by the next generation of NIR
spectrographs. The astronomical community is poised to enter a new era in NIR
spectroscopy, with multiobject instruments coming on line at multiple 8-10m
class telescopes (MOIRCS at Subaru, FLAMINGOS II at Gemini South, EMIR at the
Gran Telescopio Canarias). In this regard, the FLAMEX survey furthers one of
the original aims of the NOAO Deep Wide-Field Survey by providing a wide-area
public data set that can be used to select targets for spectroscopy.

\begin{deluxetable}{lccl}
\tighten
\tabletypesize{\scriptsize}
\tablecaption{NIR Surveys}
\tablewidth{0pt}
\tablehead{
\colhead{Survey} & \colhead{Filter}  & \colhead{Area (deg$^2$)}  & \colhead{Depth} 
}

\startdata
\citet[][2MASS]{Skrutskie1997}  &    $JHK_s$       &   All Sky   &    $K=14.5$  \\
\citet{Gardner1996}   &    $K$         & 9.84      &    $K=15.6$  \\
\citet[NDWFS \boot]{Jannuzi1999} & $K$         & 9.3     &    $K=18.6$  \\
\citet{Huang1997}     &    $K^\prime$         &  8.23     	&    $K^\prime=16$  \\
{\bf FLAMEX }	      &  \bf  $\mathbf{JK_s}$       & \bf{ \totarea }      	  &  $ \mathbf{K_s\approx19.3}$   \\
\citet[][LCIRS]{Chen2002}      &    $H$         &  1.1      &    $H\approx21.5$  \\
\citet[][MUNICS]{Drory2001} &   $JK^\prime$ &  $\sim$1      &   $K^\prime=19.5$ \\
\citet{Kummel2001}    &    $K$         &  0.91     &    $K=17$  \\
                      &                &  0.61     &    $K=17.5$  \\
\citet{Szokoly1998} &    $K_s$       &  0.6      &    $K_s=16.5$  \\
\citet[][CADIS]{Huang2001} &    $K^\prime$    & 0.2       &    $K^\prime=19.75$  \\
\citet{Daddi2000} &    $K_s$    & 0.19      &    $K_s=18.8$  \\
                      &                &  0.12     &    $K=19.2$  \\
\citet{Glazebrook1994} &    $K$       &  0.15 &        $K=17.3$  \\ 
\citet{Cristobal2003} &    $K_s$       &  $5\times10^{-2}$ &    $K_s=21$  \\  
\citet{Minezaki1998} &    $K^\prime$       & $5\times10^{-2}$    &    $K^\prime=19$  \\
                     &                     & $6\times10^{-4}$    &    $K^\prime=21$  \\
\citet{Martini2001}   &    $JHK$       & $3.4\times10^{-2}$    &    $K=18.5$  \\ 
\citet[][K20]{Cimatti2002} &  $K_s$       & $1.4\times10^{-2}$    &    $K\ga20$ \\  
\citet{McLeod1995}    &    $K$         & $5.6\times10^{-3}$ &    $K=20$  \\  
                      &                & $5.6\times10^{-4}$ &    $K=21.5$  \\  
\citet{Totani2001} &    $K^\prime$       & $1.1\times10^{-3}$     &    $K^\prime=23.5$  \\ 
\citet{Djorgovski1995}&    $K$         & $8.3\times10^{-4}$     &    $K=24$  \\ 
\citet{Moustakas1997} &    $K$         & $5.6\times10^{-4}$    &    $K\approx23.4$\tablenotemark{a}  \\ 
\citet{Bershady1998}  &    $JK$        & $4.2\times10^{-4}$    &    $K=24$  \\ 

\enddata

\tablecomments{This table is intended to be a representative rather than 
comprehensive synopsis of previous surveys. The surveys are listed in order
of decreasing survey area.  We note that different authors quote different
significance limiting magnitudes (for example $3\sigma$ vs. $5\sigma$) within
different size apertures, so the listed values provide only an approximate
comparison of survey depths.
}
\tablenotetext{a}{This survey is split between two regions with limiting
magnitudes of $K=23.1$ and $K=23.7$.}

\label{tab:surveys}

\end{deluxetable}

In this paper we present an overview of the FLAMINGOS Extragalactic
Survey. The science programs based upon the survey and details of the FLAMEX
catalogs will be presented in other papers.  The main aim in this manuscript
is to provide relevant information regarding the general survey
characteristics to the community in conjunction with the public data release,
enabling researchers interested in the NDWFS fields to assess the
applicability of this data set to their research programs.  Towards this end,
after presenting the data reductions and data quality we use the catalog in
the \boot~field to provide two brief illustrations of the science topics that
can be addressed with the FLAMEX data set.

The layout of this paper is as follows. We begin by describing the
experimental design and observations in \S\ref{sec:char}, including target sensitivities and
field geometry.  In \S\ref{sec:reductions} we present the
reduction procedure, followed by a description of the properties of
the final \boot~catalog in \S\ref{sec:catalog}.  Illustrations of initial
science and a discussion of the data release are given in
\S\ref{sec:discussion}, and the paper is summarized in
\S\ref{sec:summary}. When relevant we assume a standard cosmology with
$\Omega_M=0.27$, $\Omega_\Lambda=0.73$, and $H_0=70$ \kms Mpc$^{-1}$.

\section{Experimental Design and Observations}
\label{sec:char}

This survey is made possible by the FLorida Multi-object Imaging Near-Infrared
Grism Observational Spectrometer (FLAMINGOS), 
which when commissioned was the world's first cryogenically cooled
multi-object infrared spectrograph \citep{Elston2003}. FLAMINGOS has a fast
all-refractive optical system that can be used at telescopes slower than
f/7.5. This optical system, coupled with a 2048$\times$2048 HgCdTe HAWAII-2
array, makes FLAMINGOS a very efficient wide-field imager when used on fast,
small aperture telescopes.  For the FLAMEX survey we rely purely on FLAMINGOS'
imaging capabilities and the wide field afforded by the Kitt Peak National
Observatory 2.1m. At
this telescope FLAMINGOS has a $21\arcmin\times 21\arcmin$ field-of-view with
0.61$\arcsec$ pixels.

\subsection{Location and Field Geometry}

We initially targeted rectangular 5 \degtwo~regions within each of the two
NDWFS fields. 
The FLAMEX survey chose the NDWFS fields due to the deep optical imaging
that was in progress in these fields when the survey was begun, with the
$B_WRI$ 
data being essential for enabling derivation of accurate photometric
redshifts.  The even split between the two fields was designed to permit
follow-up observations during both observing semesters and from both
hemispheres, while still yielding coverage for large, contiguous
regions. Finally, the angular size of the survey is driven by the requirement
that we probe a sufficiently large volume to detect massive clusters at
$z>1$. For a standard $\Lambda CDM$ cosmology, one expects a surface density
of approximately $0.7-4$ deg$^{-2}$ clusters with $M>10^{14}$ M$_\odot$ at
$z>1$, depending upon the value of $\sigma_8$.\footnote{These surface densities
are calculated analytically, and are insensitive to whether one uses
the prescription from \citet{Sheth1999} or \citet{Jenkins2001}.}
 Consequently, the survey must
cover roughly $5-10$ deg$^2$ to ensure detection of a sample of massive
clusters at this epoch.

\begin{figure}
\begin{center}
\plotone{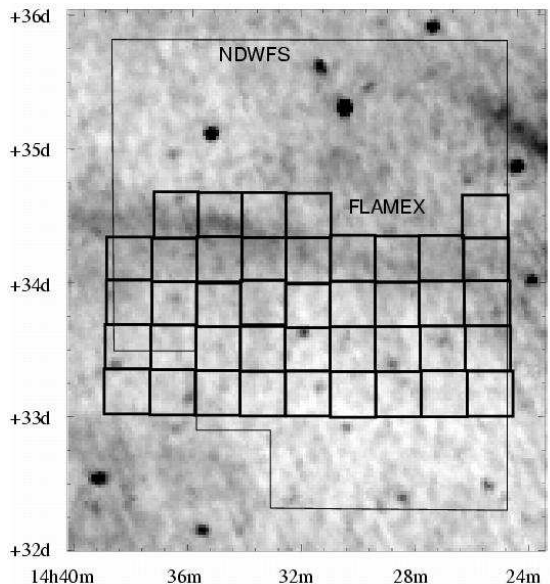}
\end{center}
\caption{Final coverage for the FLAMEX survey in the NDWFS \boot~field,
overlaid upon an IRAS 100$\mu$m image of the region \citep{Wheelock1994}. The
boxes denote individual FLAMEX pointings, with the extent of the full NDWFS
field also displayed. In the \boot~region we cover a \bootarea~\degtwo~strip.}
\label{fig:bootesfield}
\end{figure}

\begin{figure}
\epsscale{1.3}
\hskip -1.0cm\plotone{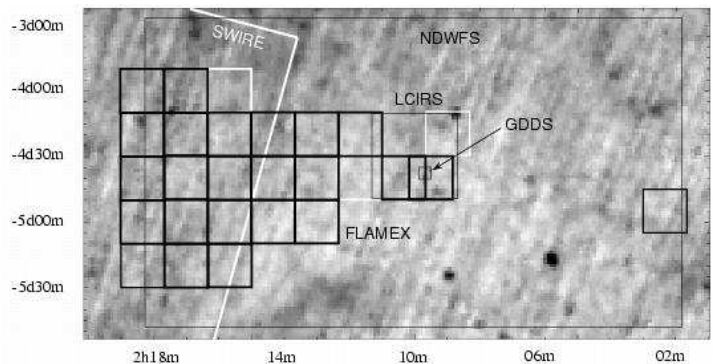}
\epsscale{1}
\caption{ Final FLAMEX coverage in the NDWFS \cetus~field. The image and boxes
are the same as in Figure \ref{fig:bootesfield}, except that three fields with
incomplete $J-$band coverage are drawn in white. For \cetus~we also overlay
the locations of the SWIRE survey (large, white box), LCIRS (large, thin black
box), and GDDS (small, thin black box). Near the GDDS field the overlap of the
FLAMEX pointings was increased to increase the depth.  The survey covers a
total of \cetarea~\degtwo~in both $J$ and $K_s$ in \cetus, with an additional
0.3 \degtwo~in $K_s$.
\label{fig:cetusfield}}
\epsscale{1}
\end{figure}

While the survey plan called for a total of 10 \degtwo, in practice worse than
expected weather (see below) limited the total coverage to
\totarea~\degtwo. In the northern (\boot) field the survey covers a
\bootarea~\degtwo~region in both $J$ and $K_s$; in the southern field (\cetus)
the corresponding coverage is \cetarea~\degtwo. The final geometry of each
field is shown in Figures \ref{fig:bootesfield}--\ref{fig:cetusfield}.  In
\cetus, this geometry was driven by following considerations. First, we
started with observations in the eastern part of the NDWFS field to best
complement the shallower $K-$band imaging from the NDWFS survey
(\citealt{Jannuzi1999}, Dey et al., in prep.), which
predominantly lies in the western part of the Cetus NDWFS field. Second, we
attempted to both maximize overlap with the \spitzer~Wide-Area Infrared
Extragalactic Survey \citep[][SWIRE]{Lonsdale2003}, which overlaps with the
eastern part of the NDWFS, and cover the Gemini Deep Deep Survey's
\citep[][GDDS]{Abraham2004} field in \cetus. The former provides valuable
additional information for photometric redshifts and constraining galaxy
populations at $z\ga1$, while the latter provides calibration for photometric
redshifts in the redshift desert ($z\approx1.5-2$).  Observations were
conducted as a gridded series of pointings, with adjacent field centers
separated by $20\arcmin$.


\subsection{Survey Specifications}

The original technical specifications for the program prescribed imaging of
the survey region in both $J$ and $K_s$ for 2 hours to reach depths of $J=22$
and $K_s=20.5$ ($5\sigma$, Vega) within 2$\arcsec$ diameter apertures, assuming
$1.5\arcsec$ FWHM seeing, which for a point source is equivalent to total
magnitudes $J=21.4$ and $K_s=19.9$. The final catalog is shallower (see \S
\ref{subsec:depth}) due to both the instrumental aberrations and worse than
expected image quality. The effective depth of the data are of course also a 
strong function of the ambient temperature (i.e. the sky brightness). We 
therefore initially imaged each field for two hours per filter, and then 
reimaged the shallower fields in later runs to obtain relatively uniform data.

\subsection{Observing Procedure}

The standard observing sequence for the survey consisted of a randomly ordered
$5\times5$ dither pattern, with 1$\arcmin$ offsets about the pointing
center. This pattern was repeated until the required total exposure time was
achieved.  Individual exposures times were dependent upon the ambient
temperature, ranging from 20--60s in $K_s$ and 90--120s in $J$.  The typical
background sky level in $K_s$ was 20,000-25,000 counts, with a maximum allowed
level of $\sim$30,000 counts. Detector nonlinearity can be corrected to better
than 1\% up to 45,000 counts; the maximum allowed level is designed to ensure
that 2MASS stars with $K_s\ge12.5$, which are used for photometric
calibration, have peak fluxes below this level.

\subsection{Observing Conditions}
\label{sec:obsconditions}

Data were obtained on the KPNO 2.1m during an allocation of 97 nights spanning
2001 December to 2004 December.  While photometric conditions were not
required, usable data were obtained on only $\sim50$\% of these nights due to
a combination of suboptimal weather conditions and telescope mechanical 
failures.
Observations were coordinated with the NOAO Survey Program "Toward a
Complete Near-Infrared Spectroscopic and Imaging Survey of Giant Molecular
Clouds" (PI: Lada). The two surveys shared nights to maximize observing
efficiency since the target fields for the two groups lie at complementary
right ascensions.

\section{Reductions}
\label{sec:reductions}

Survey imaging was reduced using the LONGLEGS pipeline (Roman et al., in prep.,
and Gonzalez et al., in prep.), which employs standard infrared reduction
procedures.  Linearity correction,\footnote{Details related to the FLAMINGOS
linearity correction and geometric distortions can be found at
\url{http://flamingos.astro.ufl.edu}.} dark subtraction, and flatfielding were
all performed in the normal fashion. A running sky, typically comprised of 8
adjacent frames, was used for sky subtraction, with objects detected in the
individual images masked before creation of the sky frames. The images were
then transformed to correct for geometric distortion, doubling the number of
pixels along both axes, and combined using a integer shift-and-add approach on these
oversampled images. The resulting stacks are astrometrically and photometrically
calibrated using the Two Micron All Sky Survey (2MASS) All Sky Data Release
\citep{Cutri2003}.
The final photometric calibration was computed using the average photometric
offset for color-selected 2MASS stars in each field, weighted by the
photometric errors. Field-to-field variations in the $J-K_s$ color of the full
stellar locus are typically at the level of 0.02 magnitudes or less,
indicating that the calibrations for the individual filters are good to this
level or better.

Because the data for the survey were obtained over an extended time period, in
many instances it was necessary to combine image stacks from different nights
(or years). When combining, we weighted the images based upon the seeing,
zeropoint, and rms noise in the input frames via the formula:
\begin{equation}
w=\frac{10^{0.4 m_0}}{\sigma (FWHM)^2},
\end{equation}
where $m_0$ is the photometric zeropoint, FWHM is the full width half-maximum
of the point spread function (PSF) at the field center, and $\sigma$ is the
rms noise in each image.
\begin{figure}
\epsscale{1.1}
\vskip -0.5cm\plotone{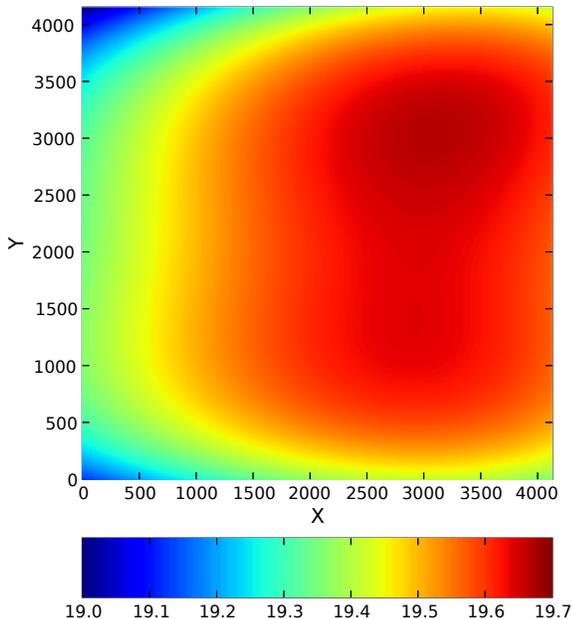}
\epsscale{1}
\caption{Map of the 50\% completeness limit for 6$\arcsec$ aperture magnitudes
for $K_s$ in one of the \boot~subfields. North is down and east is to the
left. This map is generated using simulated elliptical galaxies with sizes
consistent with the observed galaxy size distribution at $z=1-2$
\citep{Ferguson2004}. As discussed in the text, the survey depth can vary by
$\sim 0.5$ magnitudes within a given subfield.  
\label{fig:depthmap}}
\end{figure}

\section{Catalog Properties}
\label{sec:catalog}

We construct $K_s$-selected catalogs for each survey subfield using Source
Extractor version 2.3 \citep[SExtractor,][]{Bertin1996}, using dual image mode
to measure the $J$-band photometry within the same regions.  Detection is
performed in $K_s$ with a $0.76\arcsec$ FWHM gaussian convolution kernel, and we
impose a 5$\sigma$ object detection threshold. Weight maps are also used to minimize
detection of spurious. The catalogs for the individual
subfields are then merged, with strict right ascension and declination
boundaries defined between subfields to avoid double-counting of objects.  We
use completeness simulations, described below, to compute more realistic
estimates of the magnitude errors than those provided by SExtractor, since the
simulations can correctly account for correlated noise across pixels and
the impact of background fluctuations. The
typical magnitude errors from the simulations are roughly 2-3 times as large
as the SExtractor errors for most apertures, and larger for the 20$\arcsec$
aperture and the SExtactor automatic aperture (``AUTO''), which is designed to
give an estimate of the total magnitude.
\begin{figure}
\plotone{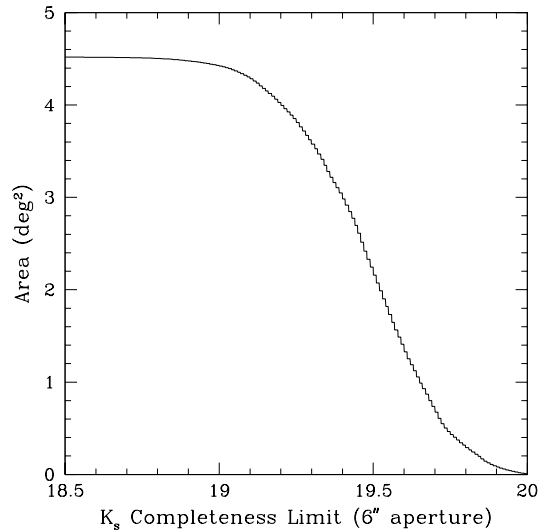}
\caption{ Survey area in \boot~as a function of the 50\% completeness limit
for the 6$\arcsec$ diameter aperture. Over 90\% of the survey region is
complete to $K_s=19.2$, with 50\% being complete to $K_s=19.5$.
Note that while the total area
covered in \boot~is \bootarea~\degtwo, the effective area after accounting for
masking and holes in the survey coverage is \netbootarea~\degtwo.
\label{fig:areavsdepth}}
\end{figure}

\subsection{Survey Depth}
\label{subsec:depth}

The depth of the survey is position-dependent due to both instrumental
aberration and shorter effective exposure times near the stack edges. The most
appropriate way to quantify this variation is via completeness simulations for
mock catalogs. We perform such simulations for each field to generate 50\%
completeness maps. Schematically, the design of the simulations is as follows.

First, we require a two-dimensional model for the point spread function for
each field. For this we take an empirical approach -- we use data from the
dense stellar fields observed by the star formation survey (see \S
\ref{sec:obsconditions}) during the same observing run.  We select the
individual frames with the best seeing, and from these data generate a lookup
table of the coordinates of bright, unsaturated stars.  Several of the runs
lack imaging from the star formation survey; for these we bootstrap off the
survey data, using stars from the survey fields with the best seeing. To test
the reliability of this bootstrap approach, we also run bootstrapped
simulations for several fields with stellar template data. In these instances
the completeness limits agree to $\la0.05$ mag in the field center and
$\la0.1-0.15$ mag in the corners.

Second, we randomly select a location in the final image to insert a fake
galaxy, and compute the corresponding location in each individual frame. We
then generate a fake galaxy, convolve it with the nearest PSF star for each
individual frame, combine these convolved model images, and add the galaxy
into the final image.
\begin{figure}
\plotone{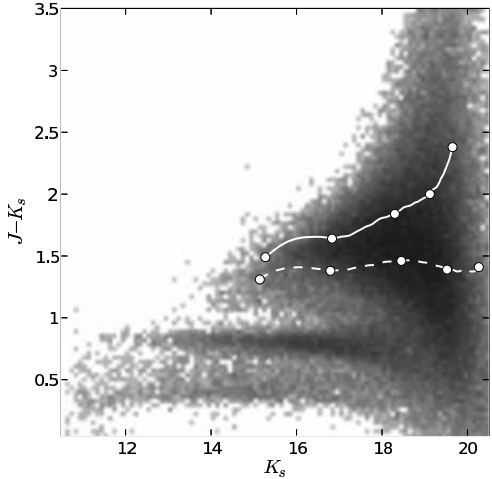}
\caption{ Infrared color-magnitude diagram for the \boot~field. The figure
includes all objects with $K_s<20.5$ that do not overlap with other objects
($\sim$84,000).  Objects are grouped in bins of width $\Delta K_s=0.1$ mag and
$\Delta J-K_s=0.035$ mag.  The greyscale runs from 0--100 objects per bin,
with a logarithmic intensity scale.  Note that the two horizontal sequences at
$J-K_s=0.4$ and $J-K_s=0.8$ correspond to Galactic stars with types later than
G5 and earlier than K5, respectively \citep{Finlator2000}. Intermediate type
stars (K5 through G5) lie between these two sequences.  The two curves are the
predicted magnitudes and colors, based upon \citet{bc2003} models with the Padova
1994 isochrones \citep{padova1994}, for $M=2\times10^{11}$ M$_\odot$ galaxies with
single burst (solid) and exponentially declining (dashed, $\tau=10$ Gyr) star
formation histories. The points correspond to $z=0.25,0.5,1,1.5,2$. }
\label{fig:cmd}
\end{figure}

For every field, in both the $J$ and $K_s$ filters, we insert 5000 fake
galaxies with magnitudes spanning the completeness limit.  The effective radii
of these galaxies are chosen to uniformly span the range of values seen at
$z=1-2$ by \citet{Ferguson2004}, who measured the distribution of half-light
radii of distant galaxies in the Hubble Deep Field--North and Chandra Deep
Field--South. Because the primary science goals of the survey focus upon the
elliptical galaxy population at $z>1$, we do not attempt to reproduce the full
morphological distribution seen at this epoch. Instead, all mock galaxies used
to calculate the completeness limits are pure bulges, with a distribution of
effective radii in the range $0.07-0.5\arcsec$ (0.5-4 kpc at $z=1$). While
obviously a simplification, in practice the bulk of the objects detected in
the survey are unresolved in the FLAMEX imaging data, making this a reasonable
approximation even for disk galaxies. To verify this assertion, we reran the
simulations in two subfields using only spiral galaxies with scale lengths
$r_d=0.6\arcsec$ (4.8 kpc at $z=1$). The derived completeness limits were
consistent to within 0.1 magnitudes with the results for pure bulges.

We use the output of the simulations to derive two-dimensional maps of the
50\% completeness limit for aperture and AUTO magnitudes measured with SExtractor.
Figure \ref{fig:depthmap} shows the derived map for a representative
subfield. As expected, the off-axis aberrations degrade the survey depth in
the corner regions.  Statistical uncertainties for these maps can be derived
via bootstrap resampling; in the image centers the completeness values have
uncertainties of $<0.03$ mag, while in the corners the uncertainties are
$\sim0.1$ mag. These values are slightly smaller that the systematic
uncertainties evidenced by comparison of simulations using the bootstrap
stellar templates with those from the star formation fields (see above).

In Figure \ref{fig:areavsdepth}, we plot the survey area as a function of the
50\% completeness depth in the \boot~region for 6$\arcsec$ diameter aperture
magnitudes, incorporating the exposure maps to reproduce the effect of masking
and dithering. More than 90\% of the survey region is complete to $K_s=19.2$,
with 50\% being complete to $K_s=19.5$.
\begin{figure}
\plotone{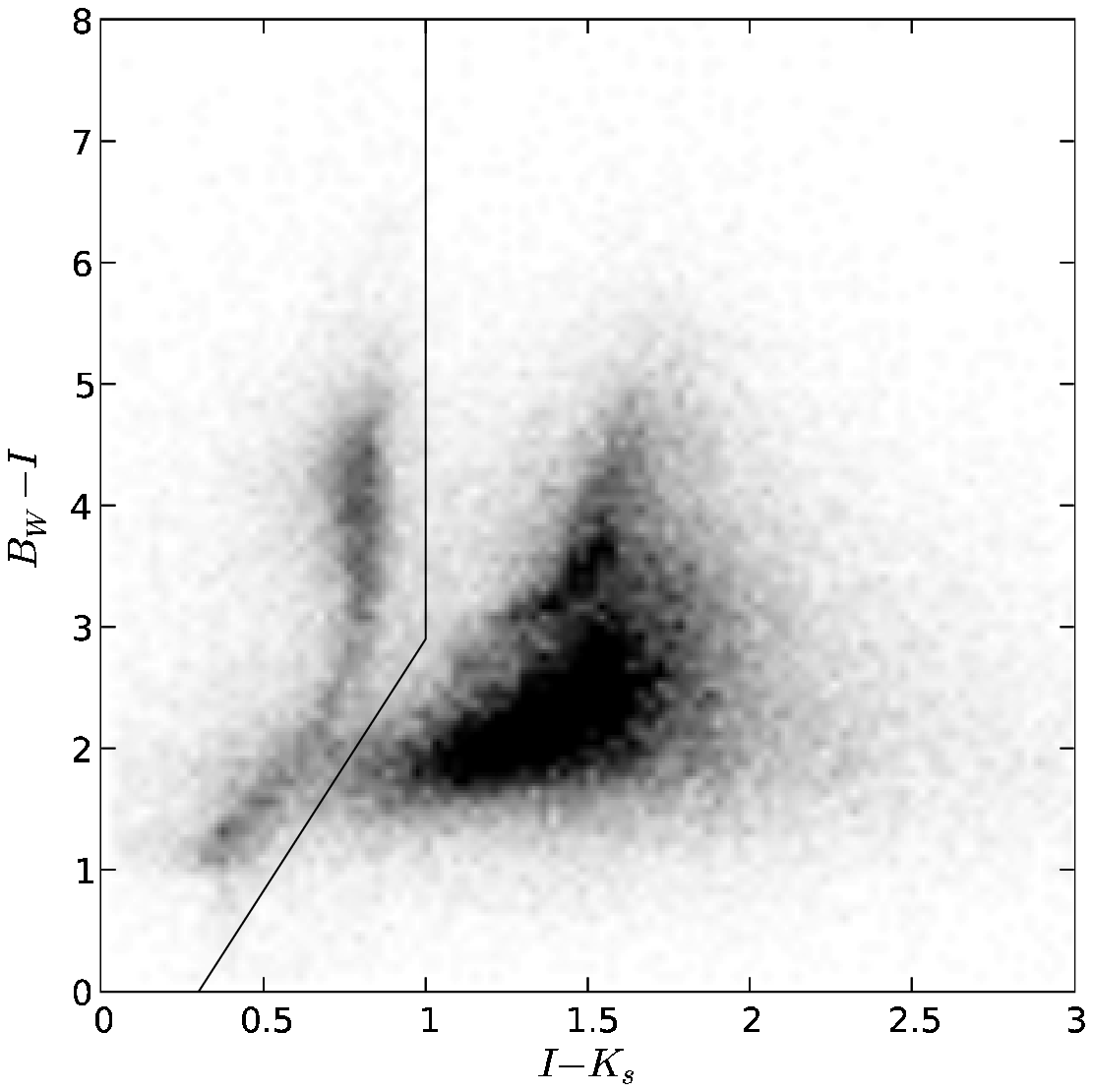}
\caption{Color-color diagram for all FLAMEX sources, using 6$\arcsec$ diameter
aperture magnitudes. The solid line denotes the division between stars and
galaxies that we utilize in \S \ref{subsec:numbercounts}.
\label{fig:bijk}}
\end{figure}

\subsection{Star-Galaxy Separation in Color Space}
\label{sec:stargal}

FLAMEX is the first deep, wide-area near-infrared survey conducted in two
filters. Our primary motivation for obtaining both $J$ and $K_s$ data was to
efficiently isolate the $z>1$ population, and these two bands, when combined
with the optical NDWFS data (and/or with \irac~data), are particularly
valuable for deriving robust photometric redshifts at $z\ga1$ (Brodwin et al.,
in prep.; Brown et al., in prep.).  The combination of near-infrared (NIR) and
optical photometry from FLAMEX and the NDWFS also find a number of other uses,
including star-galaxy separation.

The variable PSF noted above precluded use of structural information to
separate the star and galaxy populations in the FLAMEX data, but color
information is an effective substitute. The $J-K_s$ color alone provides a
simple means of performing star-galaxy separation, as can be seen in the
color-magnitude diagram (Figure \ref{fig:cmd}).  Galactic stars with types
earlier than K5 lie at $J-K_s=0.8$; later types have bluer colors, with types
later than M5 occupying a locus at $J-K_s=0.4$ \citep{Finlator2000}.
Galaxies, which occupy the redder locus of objects, are well-separated from
the stars for $z\ga0.05$.

While the $J-K_s$ color alone provides a reasonable means of performing
star-galaxy separation, inclusion of the NDWFS optical photometry can be used
to improve the separation. We have generated a matched catalog combining the
NDWFS third data release (DR3; Jannuzi et.~al., in prep) and FLAMEX data. Detections in the different bands were matched if the
centroids were within $1^{\prime\prime}$ of each other.  For extended objects,
detections in the different bands were matched if the centroid in one band was
within an ellipse defined using the second order moments of the light
distribution of the object in another band.\footnote{This ellipse was defined
with the SExtractor parameters $2 \times {\rm A\_WORLD}$, $2 \times {\rm
B\_WORLD}$, and ${\rm THETA\_WORLD}$.}  Figure \ref{fig:bijk} shows the
$B_W-I$ vs $J-K_s$ color-color diagram for the survey using 6$\arcsec$
aperture magnitudes (sufficiently large that PSF mismatch between surveys is
not an issue).  The stellar sequence can be seen to cleanly separate from the
galaxy locus, with the M stars being relatively redder in $B_W-I$ than the
bluest galaxies, and the G stars remaining bluer in $J-K_s$. There remains a
modest overlap between faint, nearby galaxies and the G star population, but
the cut shown in Figure \ref{fig:bijk} overall provides an efficient means of
separation with $\la5\%$ misidentification of galaxies at faint magnitudes
(\S\ref{subsec:numbercounts}).

\subsection{Number Counts}
\label{subsec:numbercounts}

\begin{figure}
\plotone{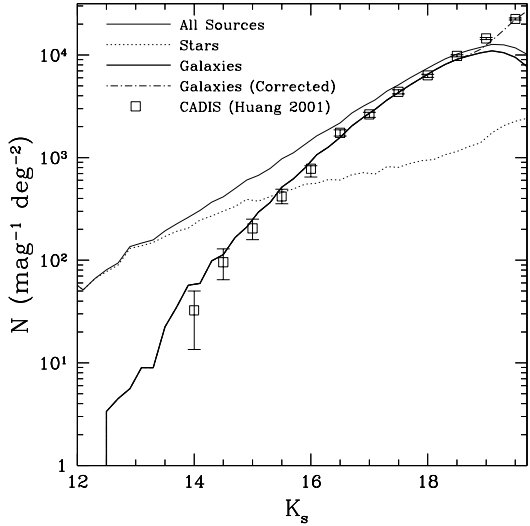}
\caption{$K_s$ differential number counts for the \boot~region. We use Source
Extractor AUTO magnitudes, which for bright extended sources agree well with
the 2MASS K20 total magnitudes. The thin solid line denotes the total number
counts in the region, while the thick solid and dotted lines correspond to the
color selected galaxies and stars, respectively. We also include the
completeness-corrected galaxy number counts (dot-dashed line) and the galaxy
number counts from the 20 arcmin$^2$ Calar Alto Deep Imaging Survey
\citep[CADIS;][]{Huang2001}, one of the largest previous $K_s$ surveys to
comparable depth. Our completeness-corrected number counts are consistent with
the CADIS values over the full range, albeit with greater statistical
precision (see also Figure \ref{fig:ncmod}).
\label{fig:numbercounts}}
\end{figure}

Galaxy number counts provide a useful means of testing the reliability of both
the star-galaxy separation (\S\ref{sec:stargal}) and the completeness
simulations (\S\ref{subsec:depth}).  In Figure \ref{fig:numbercounts} we show
the differential number counts for all sources in \boot. Using the star-galaxy
color separation, we split the number counts into the star and galaxy
contributions. The slight upturn in the stellar counts at $K_s\approx19$
indicates that we may be misclassifying up to 5\% of the galaxy population as
stars if this is not an intrinsic feature of the stellar number counts.  This
systematic should be considered for any work involving either high-precision
modelling of the number counts or statistical analysis of the local faint
galaxy population.

Figure \ref{fig:numbercounts} also provides a comparison of the FLAMEX number
counts with results from the 0.2 \degtwo Calar Alto Deep Imaging Survey
\citep[CADIS,][]{Huang2001}, one of the largest previous $K_s$ surveys to
comparable depth. Once a completeness correction is applied to the galaxy
counts (dot-dashed curve), the faint-end FLAMEX number counts are consistent
with CADIS to $K_s\ga19.5$.  This agreement at the faint end provides
validation for the completeness simulations described in \S\ref{subsec:depth}.

\begin{figure*}
\plottwo{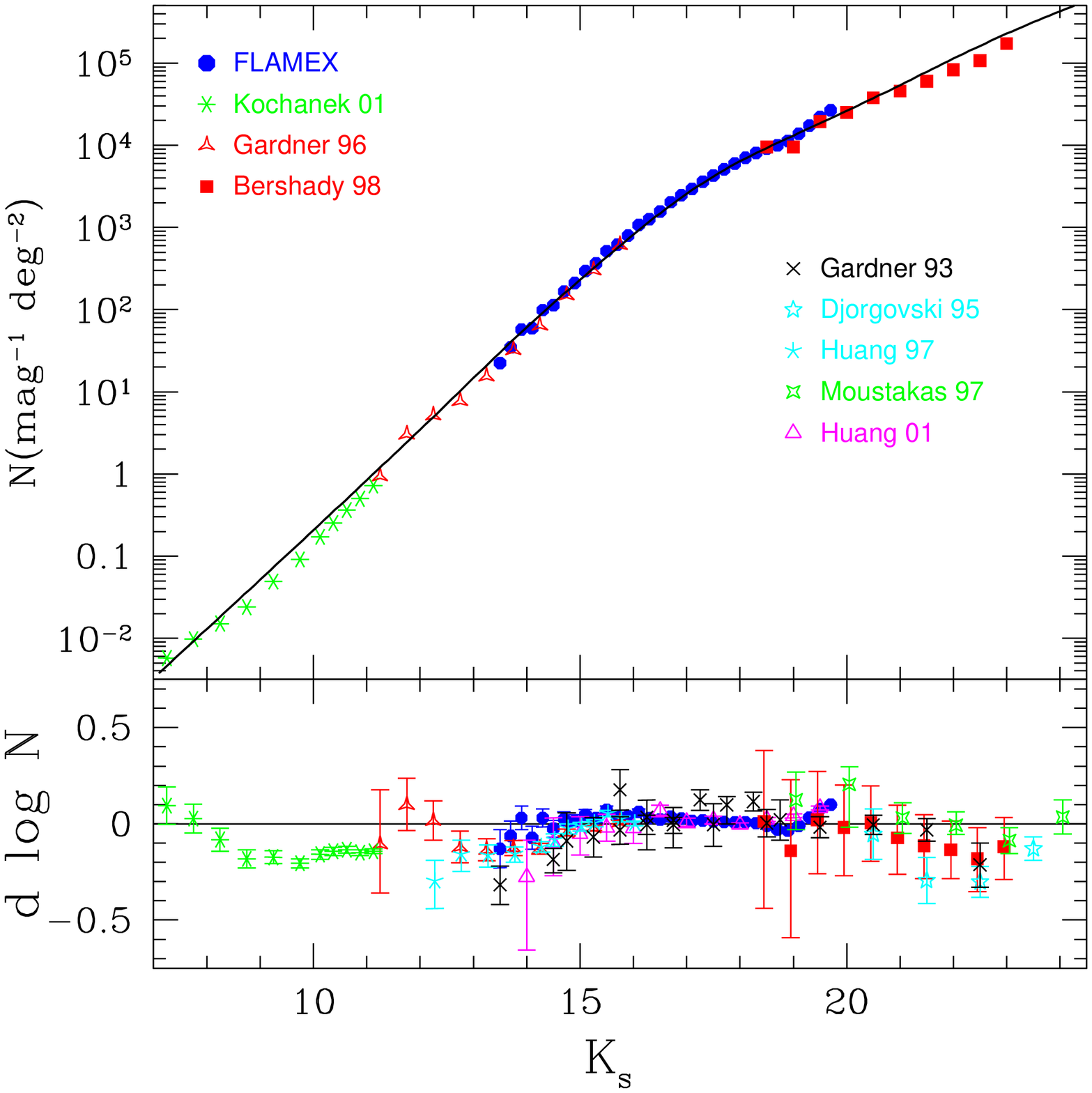}{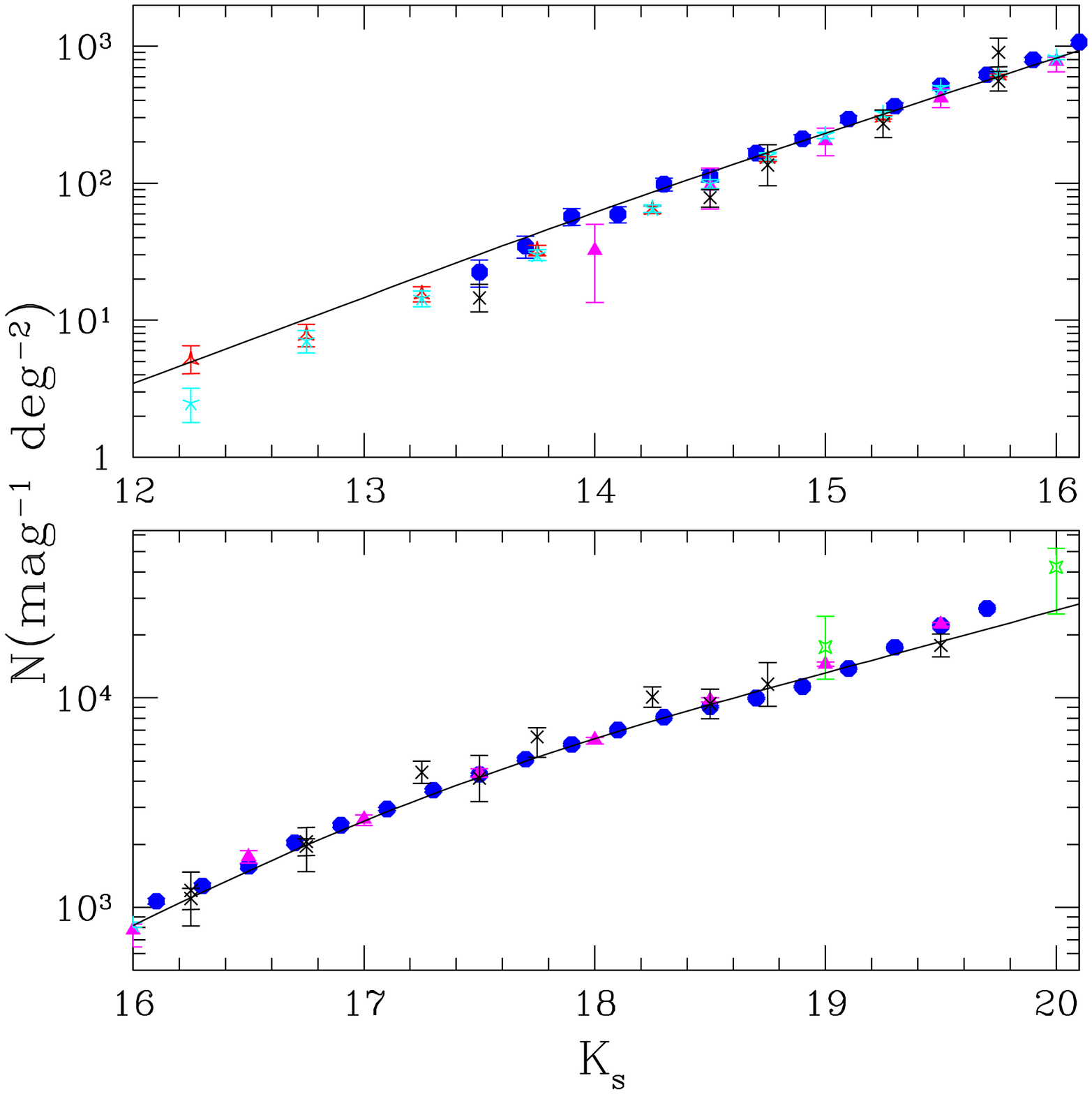}
\caption{Differential number counts from FLAMEX and a representative sample of
published data sets \citep{Gardner1993,Djorgovski1995,Gardner1996,
Huang1997,Moustakas1997,Bershady1998,Huang2001,Kochanek2001}. The left panel
shows the number counts for the full magnitude range probed by existing
studies. The FLAMEX data and a subset of the data sets are shown in the upper
part of the figure. Overlaid is a simple model constructed using {\it ncmod} 
that is similar to the one tested in \citet{Cristobal2003}. The formation 
epoch for ellipticals and spirals are set at $z=2$ and $z=1$, respectively, 
and a population of star-forming dwarfs (with a lower $\phi_*=2.8\times10^{-3} h^3$
Mpc$^{-3}$) is included to reproduce the faint end number counts seen in deep
surveys.  The lower part of the left panel shows the residuals relative to
this model, including the error bars, for all data sets. For FLAMEX the error
bars include only the Poisson uncertainty. There is good
agreement between FLAMEX and previous studies in the magnitude range probed by
FLAMEX.  To facilitate a clearer comparison, the panels on the right show the
same differential number counts, but restricted to the magnitude regime probed
by FLAMEX. The different surveys are generally consistent to within the
statistical uncertainties.
\label{fig:ncmod}}
\end{figure*}

Near-infrared galaxy number counts are also a classic diagnostic for
constraining models of galaxy evolution \citep[for
example,][]{Moustakas1997,Huang2001,Chen2002,Cristobal2003}. At these
wavelengths the K-corrections are small, and the luminosities are both less
affected by dust and less sensitive to recent star formation than optical
number counts. These factors result in samples that are closer to
mass-selected and hence simpler to model.  While the magnitude regime probed
by the FLAMEX data is not new, the large area afforded by this survey provides
significantly greater statistical precision than previous work, enabling
refined modelling.  Realistic semianalytic models, which attempt to
simultaneously match the observed counts in $K_s$ and other passbands
\citep[such as][]{nagashima2002}, are the optimal means for using this
information.  Such a detailed analysis is beyond the scope of this paper;
however, in Figure \ref{fig:ncmod} we compare with a simple classical model to
illustrate both the utility of this data set and its precision relative to
previous studies.

We use extendable galaxy number count model, {\it ncmod}, from
\citet{Gardner1998} with the same spectral templates and morphological mix of
ellipticals and spirals as \citet{Gardner1998}. Similar to
\citet{Cristobal2003}, we take the formation redshifts to be $z=2$ for
ellipticals and $z=1$ for spirals, and add a component of star-forming dwarf
galaxies with faint end slope $\alpha=-1.5$. For the general galaxy population
we use the Schechter function parameters from \citet{Kochanek2001}, while
for the star-forming dwarf population we use the same prescription as
\citet{Gardner1998} and \citet{Cristobal2003}, but with $\phi_*=3.5\times10^{-3} h_{50}^3$ Mpc$^{-3}$ 
(roughly a factor of two lower than these studies).
As found by
\citet{Cristobal2003}, this type of model qualitatively fits the number counts
over much of the observed magnitude range, albeit with some residual deviation.
Most notably, FLAMEX and the other surveys yield fewer counts than the model at the brighest
magnitudes. The FLAMEX number counts also exhibit an excess at  
the faint end ($K_s\ga 19$) 
that is statistically significant with regard to the plotted Poisson
uncertainties, albeit at a level that is small compared to the statistical 
uncertainties for most of the other samples shown. While an excess is also 
evident in the faintest CADIS data point, we caution against 
overinterpretation of this feature because this upturn occurs in the regime
where our completeness correction is significant.

\section{Discussion}
\label{sec:discussion}

The fundamental motivation for the FLAMINGOS extragalactic survey is to
facilitate systematic study of the evolution of massive galaxies and galaxy
clusters at $z=1-2$, the epoch of peak global star formation and a critical
era for both galaxy and galaxy cluster assembly. Complementing the existing
programs in the NDWFS fields at other wavelengths, FLAMEX provides rest-frame
optical photometry for the massive galaxy population at this epoch. The
combination of the FLAMEX and NDWFS data enables the robust photometric
redshifts  
and uniform selection criteria necessary for study of the galaxy and galaxy
cluster populations at this epoch. In \boot, the FLAMEX data also serves to
bridge the gap between the existing optical and \irac~data sets, facilitating
cross-association and investigation of optical drop-out populations.

The key scientific issues listed above will be the focus of a series of
separate papers.  In the following paragraphs, however, we supplement this
overview of the survey with a few illustrations of basic applications of the
FLAMEX data to provide the reader with a more quantitative sense of the
science that can be achieved.  In particular, we provide a brief analysis of
the extremely red object (ERO) population and a synopsis of our galaxy cluster
search before discussing our public data release.  These items are intended
simply to highlight a few uses of the FLAMEX data set -- examples of other,
multiwavelength programs (either underway or planned) that use FLAMEX include
a high-redshift quasar search, investigation of the morphology-density
relation at $z>1$, construction of empirical optical to mid-IR galaxy spectral
energy distributions, and investigation of the ensemble properties of distant
galaxy clusters.

\begin{figure*}
\plottwo{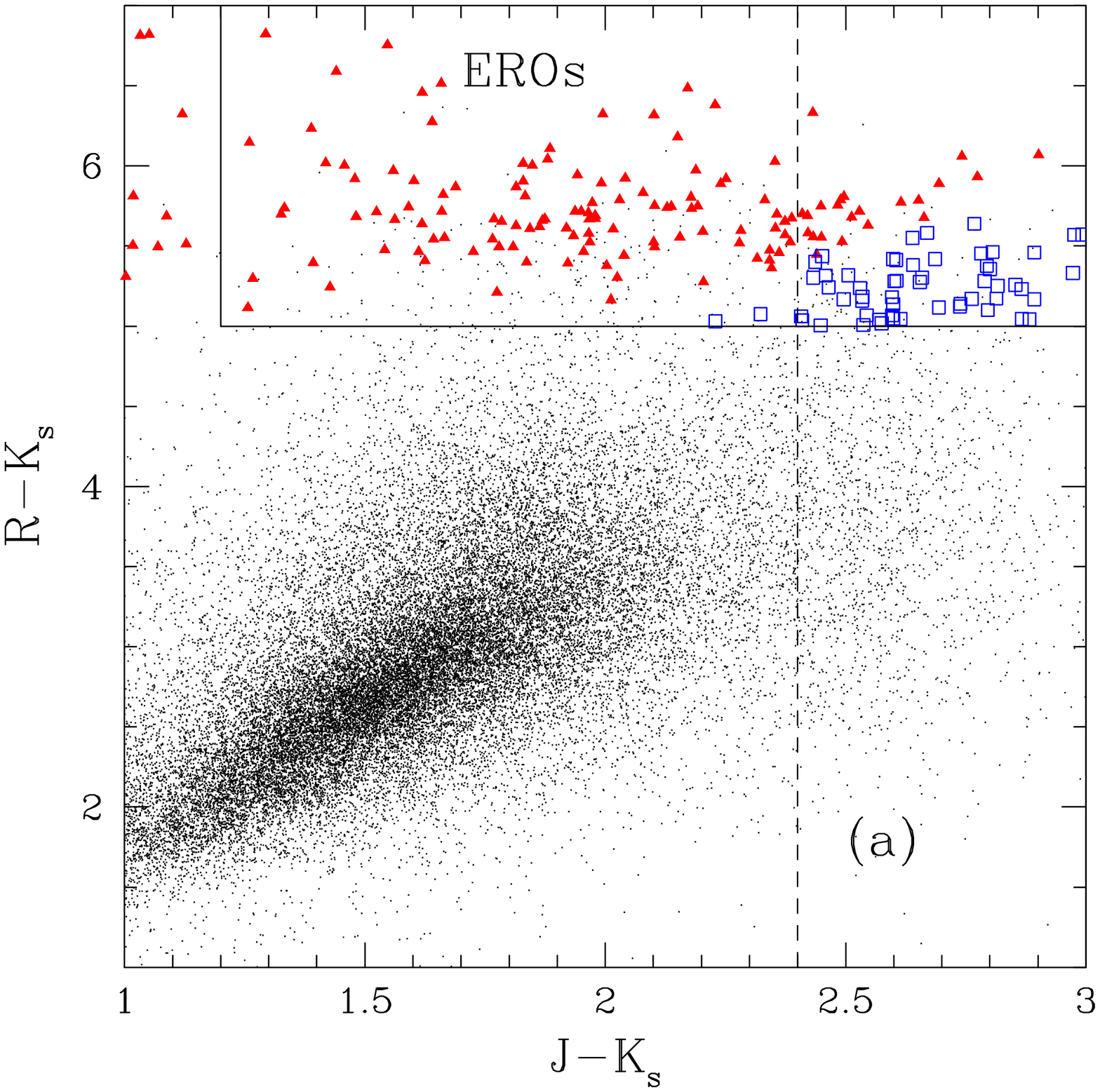}{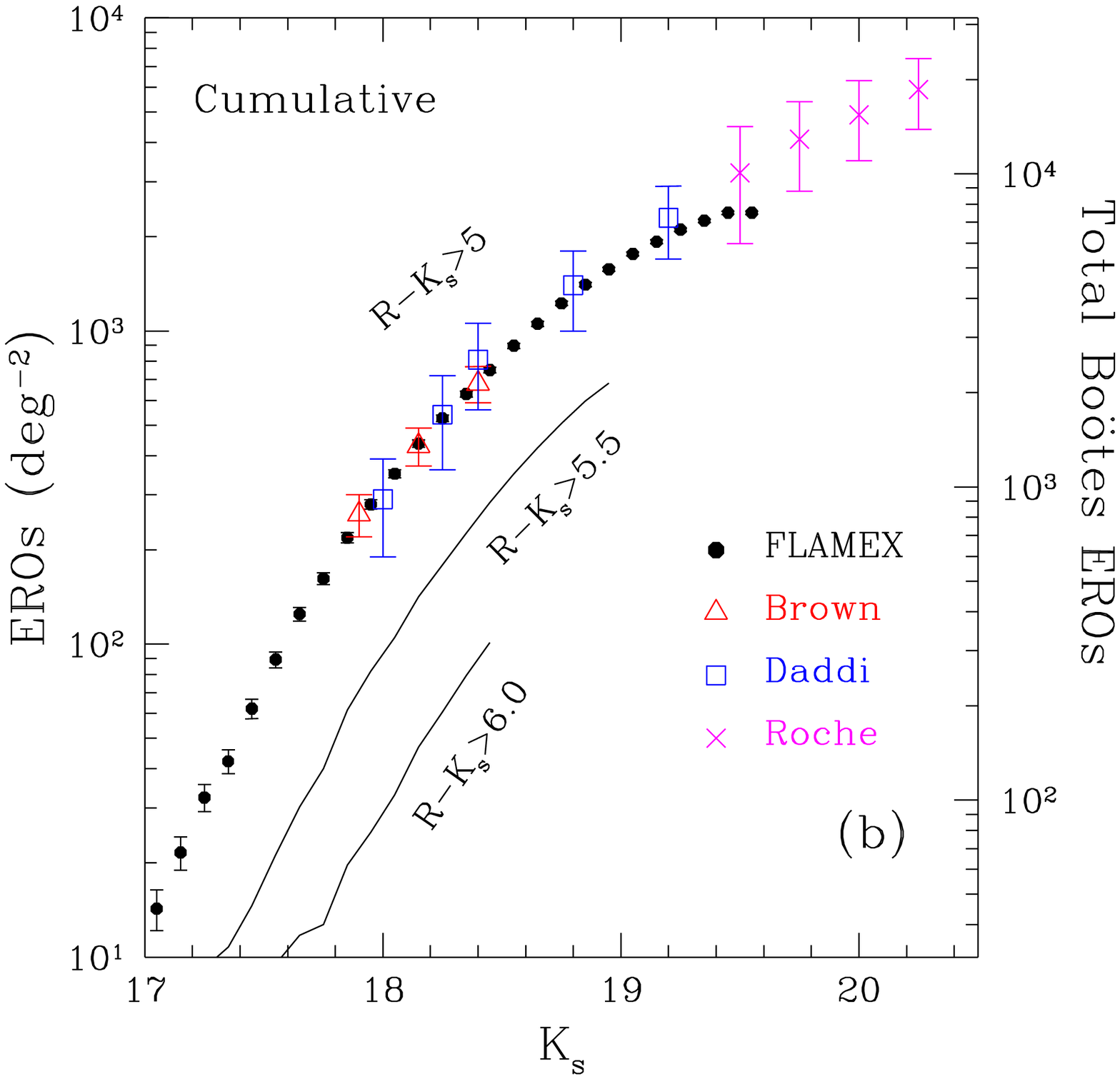}
\caption{($(a)$ $R-K_s$ versus $J-K_s$ color distribution of FLAMEX sources in
\boot~for a subset of sources with $17<K_s<19$.  EROs are identified as
objects with $J-K_s>1.2$ and $R-K_s>5$, with all colors calculated using
$6\arcsec$ aperture magnitudes. Filled triangles in this figure correspond to
objects with $R>24.5$ (roughly equivalent to an $R-$band SNR=4 in a $6\arcsec$
aperture), and should be considered lower limits on the $R-K_s$ color. Roughly
40\% of the EROS have lower limits in $R-K_s$.  Open 
boxes correspond to objects that qualify as EROs from their $R-K_s$ color, but 
have only lower limits on the $J-K_s$ color.  The vertical dashed line indicates that 50\% of sources 
with  $J-K_s>2.4$ are undetected in the $J-$band magnitude.
$(b)$ Cumulative surface density of EROs detected within a 4.2 deg$^2$ subregion
of the \boot~field.  For
$R-K_s>5$ we present the FLAMEX \boot~data as filled circles and compare the
observed number density with the results of several previous studies,
including \citet{Daddi2001}, \citet{Roche2002}, and \citet{Brown2005}.  The
FLAMEX counts for redder cuts ($R-K_s>5.5$ and $R-K_s>6$) are presented as
solid lines down to limiting magnitudes corresponding to $R=24.5$. We note
that these curves are both roughly parallel to the $R-K_s>5$ data.  The total
number of EROs in the \boot~field is shown on the right axis, demonstrating
that the sample includes $>7,000$ EROs with $K_s<19.5$.  We apply no
completeness correction to the FLAMEX data, and hence the true surface density
exceeds the FLAMEX points for $K_s\ga19$ (see Figure \ref{fig:numbercounts}).
There is also no correction applied to convert some of the other surveys from
$K$ to $K_s$, but for EROs the difference is minimal. 
 \label{fig:nm_eros}}
\end{figure*}

\subsection{EROs}

First discovered by \citet{Elston1988}, extremely red objects (EROs) are
traditionally defined as objects with $R-K_s>5$ \citep[although a range of
definitions exist in the literature; see][]{McCarthy2004}.  While initially
speculated to correspond to a population of protogalaxies at $z>6$, the ERO
population is now understood to be predominantly comprised of two distinct types
of galaxies -- dusty starbursts and galaxies dominated by old stellar
populations, both at $z\sim0.8-2$ \citep[e.g.,][]{Elston1989,Cimatti2002,Yan2004}.  For the dusty
starbursts, strong extinction in the rest-frame near-ultraviolet drives the
observed color; for the old galaxies, the red color is a simple consequence of
the 4000 \AA~break passing redward of $R-$band.  These two populations
together are believed to constitute the dominant progenitor population for
present-day elliptical galaxies, a belief that is supported by their space
densities \citep{Daddi2000} and strong spatial clustering \citep[][ but see
\citealt{Yan2004} for concerns with this interpretation]{Daddi2000,Brown2005}.
However, as a consequence of the strong spatial clustering and relatively low
surface density of EROs, large area surveys are required for precise
statistical analysis of the population.

The FLAMEX survey provides an order of magnitude increase in EROs sample size
compared to other recent studies.  We define EROs as being objects with
$R-K_s>5$ within a $6\arcsec$ aperture, imposing the additional requirement
that $J-K_s>1.2$ to eliminate potential contamination from stellar sources
with spurious optical photometry (Figure \ref{fig:nm_eros}$a$).  By this
definition, in the \boot~field alone we have over 7,000 EROs at $K_s<19.5$
(Figure \ref{fig:nm_eros}$b$); for comparison, the largest sample used thus
far to study the clustering of EROs contained 671 objects down to $K=18.4$
within a 1.4 $\deg^2$ region in the NDWFS \boot~field
\citep{Brown2005}.  The FLAMEX counts are consistent with previous work, with
the advantage that the large field enables identification of the brightest,
rarest EROs and measurement of the number counts as bright at $K_s\approx17$.
This sample is also sufficiently large to begin to quantify the number density
distribution as a joint function of magnitude and color (see the curves for
redder cuts in Figure \ref{fig:nm_eros}$b$), providing a more stringent
observational test for semianalytic evolutionary models.
\begin{figure*}
\epsscale{0.45}
\plotone{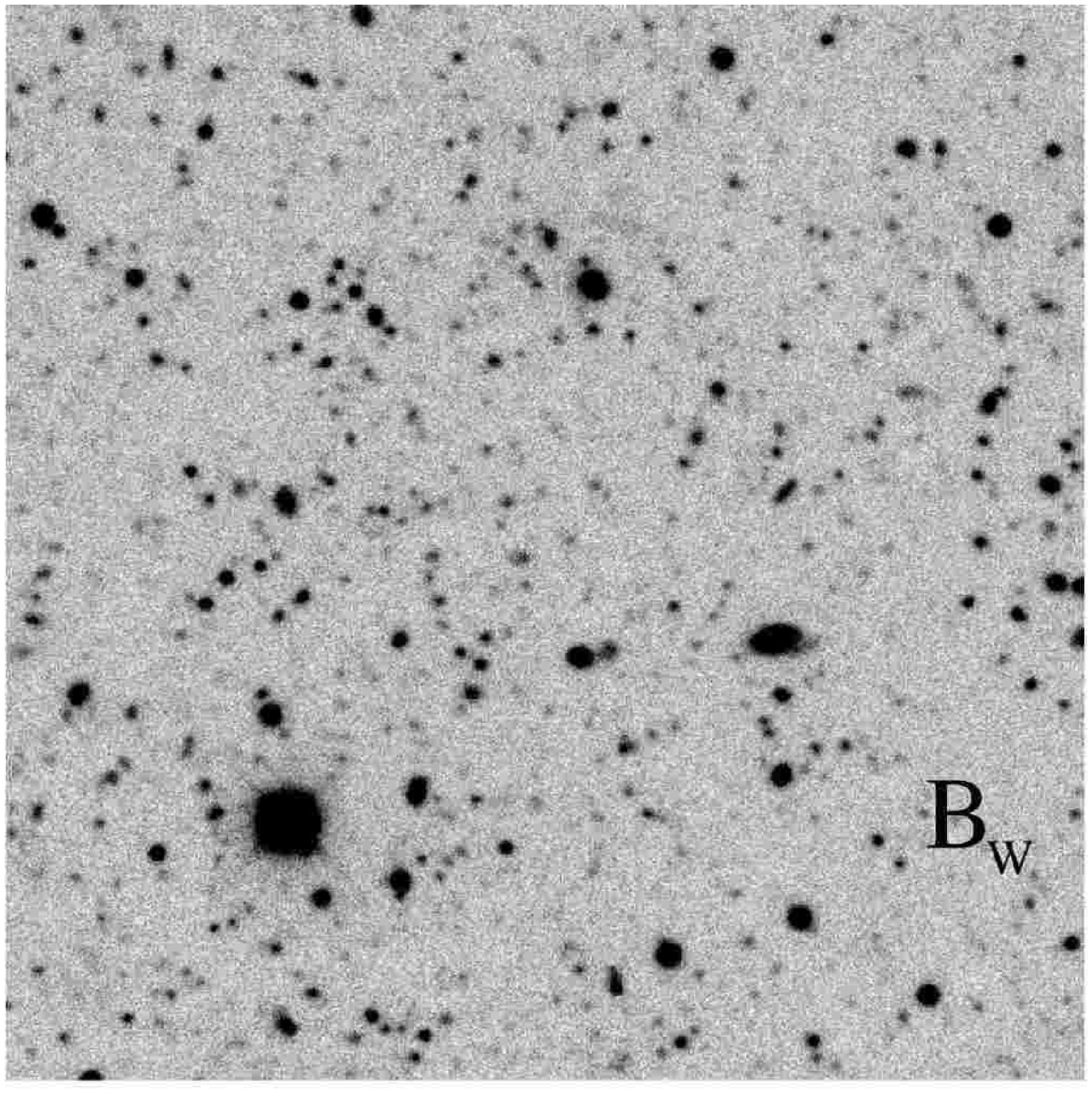}\plotone{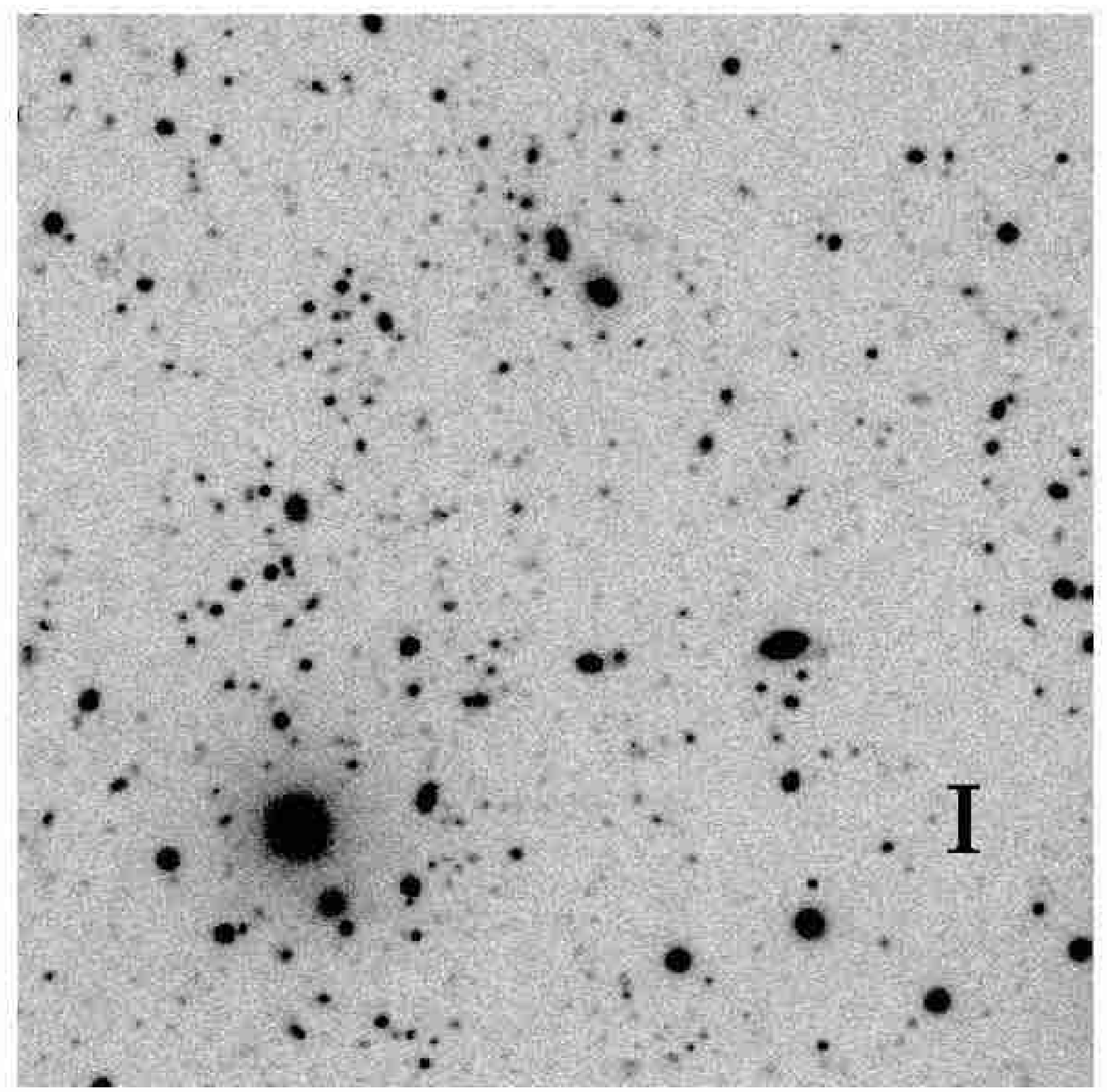} \plotone{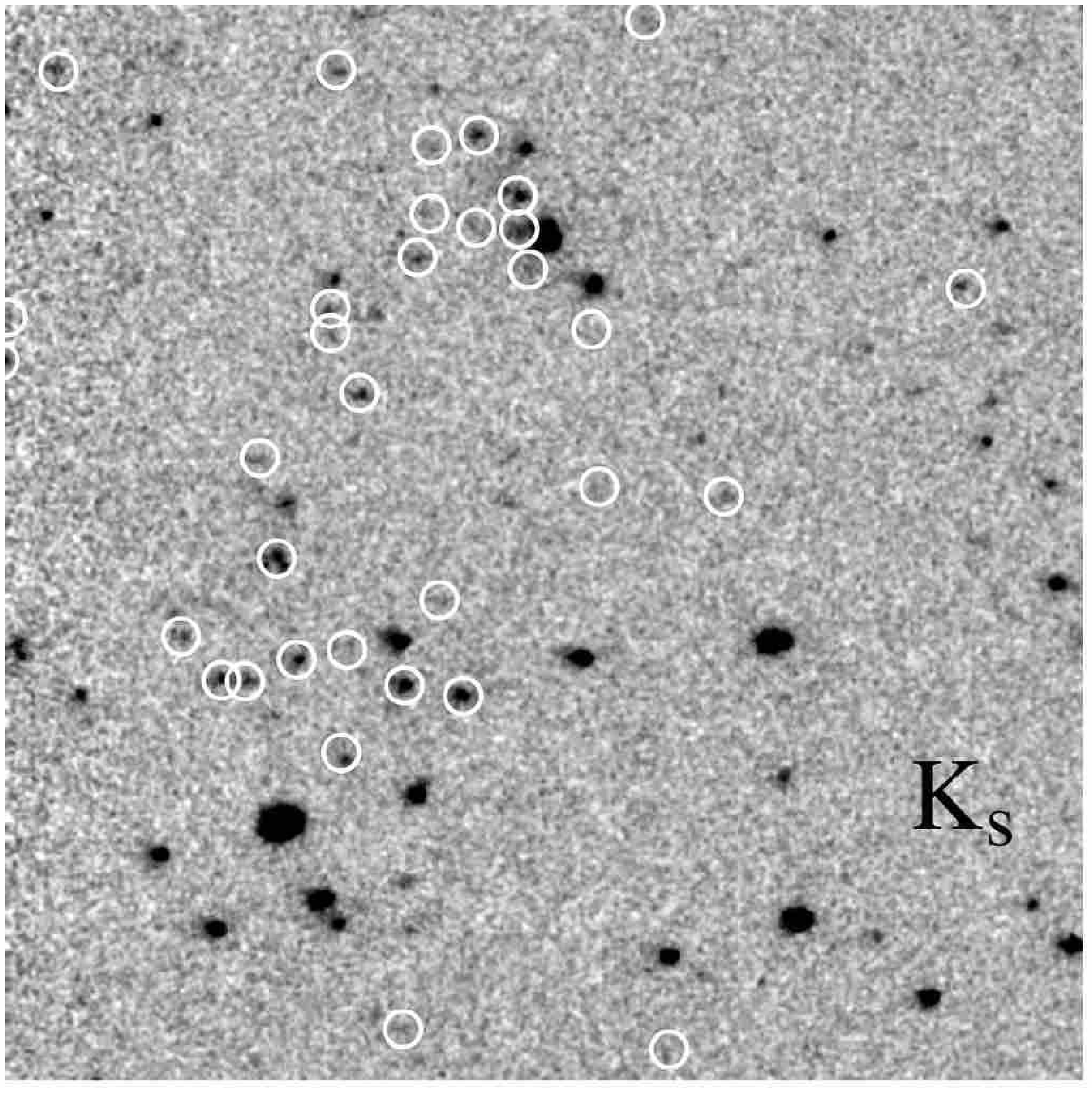}\plotone{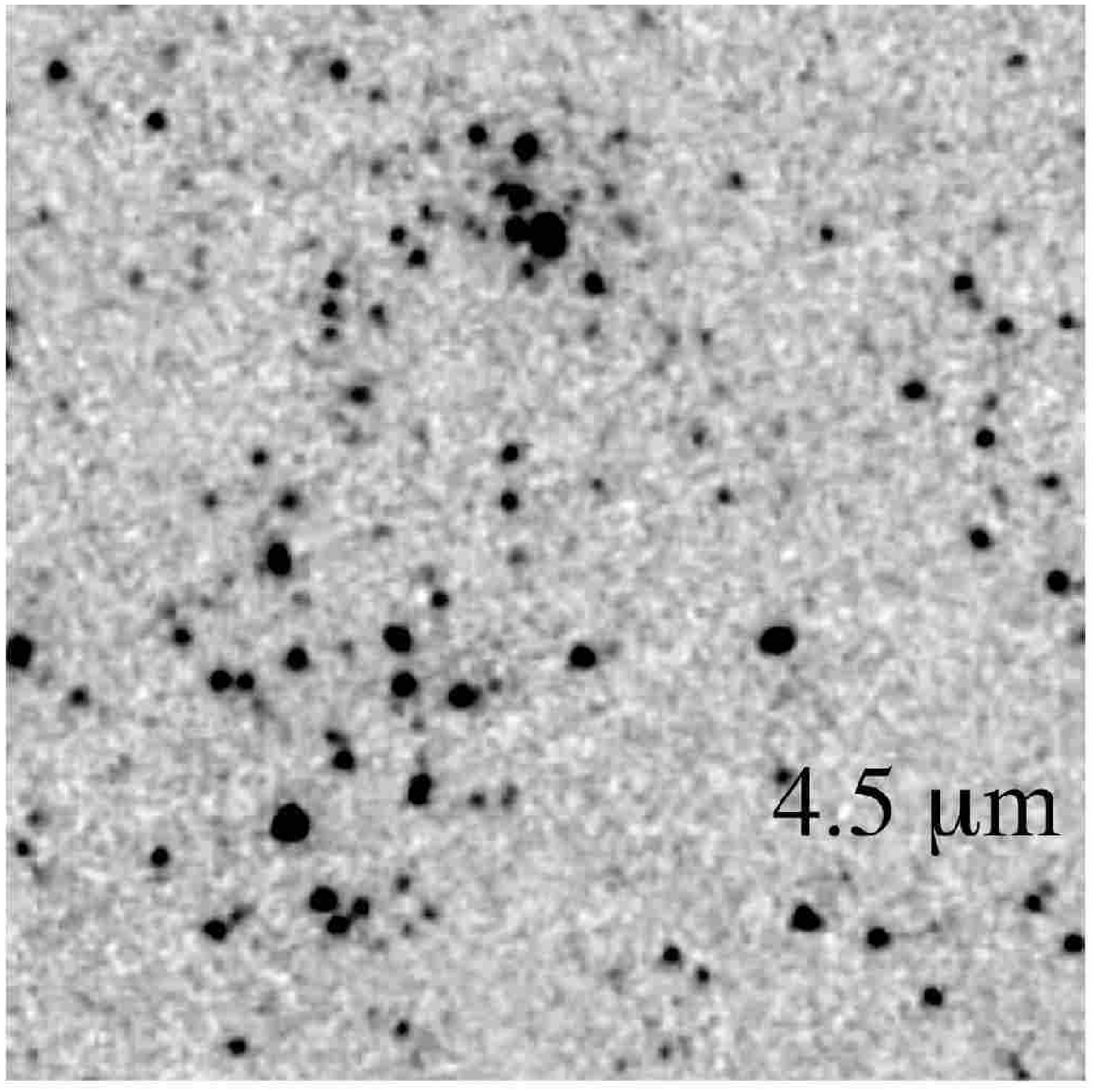}
\caption{The four image panels, each $3\arcmin\times3\arcmin$, show the
core region of CL1432.5+3332.8 in $B_W$, $I$,  $K_s$, and IRAC 4.5$\mu m$.
The cluster is visible as an excess of red galaxies in the $K_s$ and 
4.5$\mu$m bands, and in the $K_s-$band image we overlay circles to denote all
galaxies with photometric redshifts $1\le z\le 1.2$. The centroid of the 
cluster detection is located at $\alpha=14$h32m29.2s, $\delta=+33$d32m48s 
(J2000), which is near the dense clump of red galaxies in the upper, central 
part of the images.
\label{fig:cl_7.1_color.eps}}
\end{figure*}

\subsection{Galaxy Clusters}

One of the principal motivations for FLAMEX is identification of a large
sample of galaxy clusters at $z>1$.  In the standard paradigm of hierarchical
structure formation, we expect the formation of the first massive galaxy
clusters to occur at at $z=1-2$. The comoving cluster number density at
$M>10^{14}$ M$_\odot$ increases by roughly a factor of 30 between $z=2$ and
$z=1$ (WMAP cosmology), with the precise number density and degree of
evolution strongly sensitive to the details of the underlying cosmological
model (particularly $\sigma_8$ and $\Omega_M$).  Aside from the cosmological
applications \citep[see for example][]{Haiman2001,Borgani2001,WangS2004},
detailed studies of homogeneous cluster samples at this epoch are essential
for understanding the evolution of the cluster galaxy population and how
the interplay between the galaxies and host clusters impacts this evolution.
Observations at this epoch constrain the evolution of the halo occupation
distribution (HOD), the formation epoch of cluster ellipticals, evolution of
the color-magnitude relation, and the morphological evolution of cluster
galaxies.
\begin{figure}
\begin{center}
\rotatebox{270}{
\epsscale{0.8}
\plotone{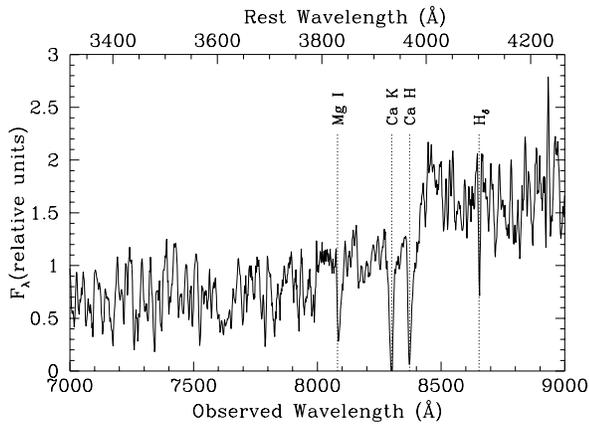}
}
\end{center}
\caption{Spectrum of a galaxy at $z=1.110$ in
CL1432.5+3332.8, after smoothing with a 5 pixel boxcar filter. Several
absorption features common to early-type galaxies are marked.
\label{fig:spectrum}}
\end{figure}
\begin{figure}
\epsscale{1}
\plotone{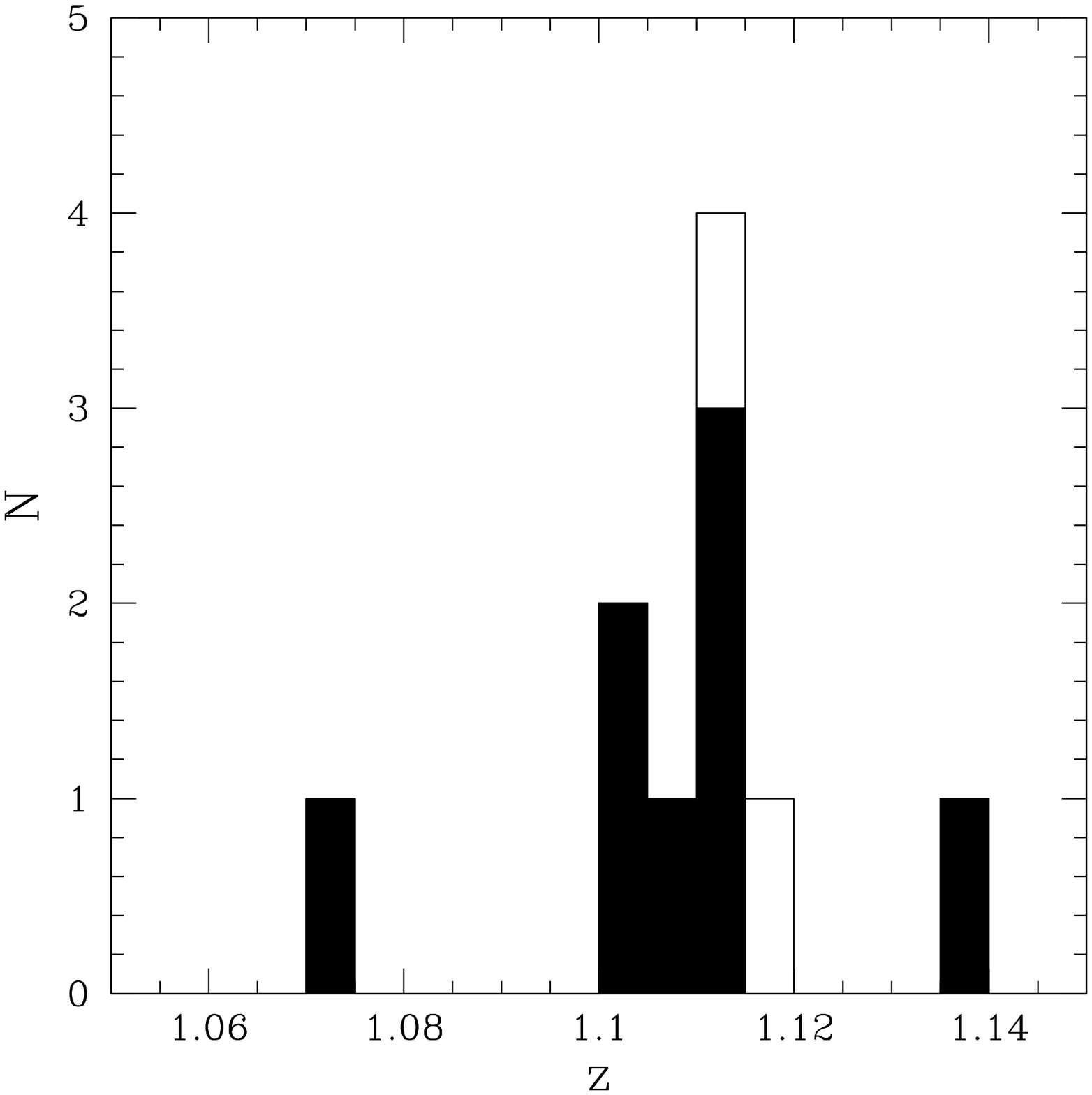}
\caption{Redshift distribution from Keck observations of CL1432.5+3332.8.
The solid histogram includes only galaxies with photometric redshifts
$0.95<z\le1.25$, while the open histogram includes all spectroscopic sources.
The derived redshift and dispersion are $<z>=1.109$ and $\sigma=920\pm230$
\kms.
\label{fig:zdist}}
\end{figure}

Infrared imaging, coupled with shorter wavelength photometry, provides an
efficient means of detecting clusters at this epoch. The infrared data enable
construction of roughly stellar mass-selected galaxy samples, and the
combination of optical and infrared data enables derivation of photometric
redshifts. The original plan for this program was to use the joint FLAMEX and
NDWFS data sets to derive photometric redshifts, and then detect galaxy
overdensities within redshift slices. We developed a detection technique
within this framework that is generically applicable to any photometric
redshift catalog. The photometric redshift catalog is used to subdivide the
data set into overlapping redshift bins, and we then employ a wavelet
algorithm with bootstrap resampling to identify statistically significant
overdensities on a fixed physical scale (a complete description will be
provided in Gonzalez et al., in prep.).

Over the course of the survey this original plan has evolved in one
significant regard.  Joining forces with the IRAC Shallow Survey
\citep{Eisenhardt2004}, we are now using the combination of NDWFS, FLAMEX, and
IRAC (3.6$\mu$m and 4.5$\mu$m) photometry to derive more robust photometric
redshifts (Brodwin et al., in prep.) and probe more deeply than would be
possible with either the NDWFS and FLAMEX or NDWFS and IRAC data sets
alone. The IRAC data are deeper than FLAMEX, allowing us to push further down
the cluster luminosity function, while the FLAMEX data significantly reduce
the scatter in photometric redshifts at $z\ga1.4$ (Brodwin et al., in prep.).
We also now fold in the full redshift probability distribution for each
source, which minimizes the noise from galaxies with poorly constrained
redshifts.  Additionally, our joint team is using the same technique in a
complementary survey to detect clusters within the full 9.3 $\deg^2$ NDWFS
\boot~field using only the NDWFS and IRAC data (Eisenhardt et al., in prep.).

In this paper we present one of the first spectroscopically confirmed $z>1$
clusters detected with the combined NDWFS, FLAMEX, and IRAC data sets,
CL1432.5+3332.8 at $z=1.11$ (Figure \ref{fig:cl_7.1_color.eps}).  To
identify clusters within the FLAMEX region we perform a wavelet search using
the photometric redshift probability distributions derived from the combined
data sets, with the wavelet scale size defined to detect structures with
physical scales $r\approx200-800$ kpc. Monte Carlo simulations are then used
to quantify the significance of each detection. This search identifies the
cluster CL1432.5+3332.8 as one of the three highest significance
overdensities at $z>1$, with an estimated redshift $z=1.2$ (Figure
\ref{fig:cl_7.1_color.eps}).

Spectroscopic confirmation of this candidate was obtained on 3 June 2005 UT
using LRIS on Keck 1.  We used a single slit mask for the observations which
were carried out in clear conditions with 0.9 arcsec seeing.  Five 1800s
exposures were obtained, with the target galaxies shifted within the slitlets
between each exposure.  The data were reduced using standard long slit
techniques \citep[e.g.][]{Stanford1997}.  The resulting redshifts include 8
galaxies at $1.10 < z < 1.12$. Of these, five exhibit absorption line features
indicative of old stellar populations (Figure \ref{fig:spectrum}), arguing
that we are observing a cluster with an established early-type population.  A
histogram of the redshifts is shown in Figure \ref{fig:zdist}.  The resulting
velocity dispersion is $\sigma = 920\pm230$ \kms, as calculated using the
gapper method (Beers, Flynn, \& Gebhardt 1990), confirming that
CL1432.5+3332.8 is a massive galaxy cluster.

The cluster CL1432.5+3332.8 is presented here as a proof of concept to
demonstrate that we can successfully identify clusters at $z>1$ using the
combined \boot~data sets.  A single cluster detection of course cannot
constrain the fidelity or completeness of catalogs derived with this data
set. These are issues that will be addressed in subsequent papers, and in
particular we refer the reader to \citet{Stanford2005}, Brodwin et al. (in
prep.) and Eisenhardt et al. (in prep.) for further explorations of the
photometric redshifts and cluster catalogs derived from the NDWFS+FLAMEX+IRAC
and NDWFS+IRAC data sets.

\subsection{Public Catalogs}
\label{sec:pubcat}

The FLAMINGOS Extragalactic Survey received time through the NOAO Large Survey
Program and provides public access to the catalogs for the full FLAMEX
region. The most recent version of the catalogs, accompanied by a detailed
description, is posted on the web (starting September 30, 2005) at the
following mirrors: \url{http://flamingos.astro.ufl.edu/extragalactic/} and
\url{http://www.noao.edu/noao/noaodeep/}.

\section{Summary}
\label{sec:summary}

We present an overview of the FLAMINGOS Extragalactic Survey (FLAMEX), a
near-infrared imaging survey conducted at the KPNO 2.1m with the FLAMINGOS
instrument.  FLAMEX is the first deep, wide-area NIR survey to image in both
the $J$ and $K_s$ filters, and is larger than any previous surveys to
comparable depth.  Located within the NOAO Deep-Wide Field Survey regions, FLAMEX is
designed to complement the NDWFS optical photometry and other existing
multiwavelength public data sets, with the intent that it be a resource to the
astronomical community. We therefore make the full catalogs from this 
program publicly available via the web pages listed in \S \ref{sec:pubcat}.

The survey area (\totarea~\degtwo) is divided between the northern (\boot) and
southern (\cetus) NDWFS regions, with \bootarea~\degtwo~in \boot~and
\cetarea~\degtwo~in \cetus. We provide an initial characterization of the
data set based upon the $\sim150,000$ sources detected (5$\sigma$) within the
\boot~field. The data in this region reach a median depth of $K_s\sim19.3$
(6$\arcsec$ aperture), with a positionally dependent limiting magnitude that
we model using extensive Monte Carlo simulations. Using the \boot~catalog in
conjunction with optical NDWFS photometry we are able to perform a color-based
star-galaxy separation that is robust for all sources except faint, blue
galaxies at $z\sim0$. We demonstrate that the \boot~number counts are fully
consistent with results from previous, smaller area surveys down to
$K_s=19.6$, but with vastly reduced statistical uncertainty. We discuss the
utility of this data set for quantifying the number density and clustering
properties of the ERO population. The FLAMEX \boot~catalog alone contains an
order of magnitude more EROs than any published studies. Finally, we describe
one of the first spectroscopically confirmed clusters at $z>1.1$, which was
identified using photometric redshifts derived from the joint FLAMEX, NDWFS,
and \irac~Shallow Survey catalogs.

\acknowledgements

The authors are deeply grateful to Elizabeth Lada for her many contributions 
to FLAMINGOS and our survey. In particular we thank her for critical 
assistance, advice, and guidance during the transfer of the survey 
leadership. Her personal and professional support were key to the
success of the program.  We also express our sincere thanks to the NOAO 
Survey TACs and NOAO 
for granting time for this survey through the NOAO Survey Progam, and to Ron 
Probst and the observatory staff at Kitt Peak for support of the FLAMINGOS 
instrument.   Additionally, we appreciate the efforts of the full FLAMINGOS 
instrument 
team at the University of Florida, including Jeff Julian, Kevin Hanna, and 
David Hon, and are grateful to Jonathan Gardner for his suggestions as
referee.
RJE acknowledges support from the National Science Foundation via a PECASE
grant (AST-9875448). AHG acknowledges support from the NSF Astronomy and
Astrophysics Postdoctoral Fellowship program under award AST-0407085 and from
an NSF Small Grant for Exploratory Research under award AST-0436681.  The work
of MB, PRME, and DS was carried out at the Jet Propulsion Laboratory,
California Institute of Technology, under a contract with NASA. JJM acknowledges 
support from the NASA Long Term Astrophysics award NAG 5-11415.

This publication makes use of data products from the Two Micron All Sky
Survey, which is a joint project of the University of Massachusetts and the
Infrared Processing and Analysis Center/California Institute of Technology,
funded by the National Aeronautics and Space Administration and the National
Science Foundation.  This work is based in part on observations made with the
\spitzer~{\it Space Telescope}, which is operated by the Jet Propulsion Laboratory,
California Institute of Technology, under a contract with NASA.

\vskip 0.1cm
Facilities: KPNO 2.1m (FLAMINGOS), Keck 1 (LRIS), Spitzer (IRAC), KPNO 4m (MOSAIC-1)

\bibliographystyle{apj}
\bibliography{ms}

\begin{thebibliography}{69}
\expandafter\ifx\csname natexlab\endcsname\relax\def\natexlab#1{#1}\fi

\bibitem[{{Abraham} {et~al.}(2004){Abraham}, {Glazebrook}, {McCarthy},
  {Crampton}, {Murowinski}, {J{\o}rgensen}, {Roth}, {Hook}, {Savaglio}, {Chen},
  {Marzke}, \& {Carlberg}}]{Abraham2004}
{Abraham}, R.~G. {et~al.} 2004, \aj, 127, 2455

\bibitem[{{Adelberger} {et~al.}(2005{\natexlab{a}}){Adelberger}, {Erb},
  {Steidel}, {Reddy}, {Pettini}, \& {Shapley}}]{Adelberger2005a}
{Adelberger}, K.~L., {Erb}, D.~K., {Steidel}, C.~C., {Reddy}, N.~A., {Pettini},
  M., \& {Shapley}, A.~E. 2005{\natexlab{a}}, \apjl, 620, L75

\bibitem[{{Adelberger} {et~al.}(2005{\natexlab{b}}){Adelberger}, {Steidel},
  {Pettini}, {Shapley}, {Reddy}, \& {Erb}}]{Adelberger2005b}
{Adelberger}, K.~L., {Steidel}, C.~C., {Pettini}, M., {Shapley}, A.~E.,
  {Reddy}, N.~A., \& {Erb}, D.~K. 2005{\natexlab{b}}, \apj, 619, 697

\bibitem[{{Adelberger} {et~al.}(2003){Adelberger}, {Steidel}, {Shapley}, \&
  {Pettini}}]{Adelberger2003}
{Adelberger}, K.~L., {Steidel}, C.~C., {Shapley}, A.~E., \& {Pettini}, M. 2003,
  \apj, 584, 45

\bibitem[{{Bershady} {et~al.}(1998){Bershady}, {Lowenthal}, \&
  {Koo}}]{Bershady1998}
{Bershady}, M.~A., {Lowenthal}, J.~D., \& {Koo}, D.~C. 1998, \apj, 505, 50

\bibitem[{{Bertelli} {et~al.}(1994){Bertelli}, {Bressan}, {Chiosi}, {Fagotto},
  \& {Nasi}}]{padova1994}
{Bertelli}, G., {Bressan}, A., {Chiosi}, C., {Fagotto}, F., \& {Nasi}, E. 1994,
  \aaps, 106, 275

\bibitem[{{Bertin} \& {Arnouts}(1996)}]{Bertin1996}
{Bertin}, E., \& {Arnouts}, S. 1996, \aaps, 117, 393

\bibitem[{{Borgani} {et~al.}(2001){Borgani}, {Rosati}, {Tozzi}, {Stanford},
  {Eisenhardt}, {Lidman}, {Holden}, {Della Ceca}, {Norman}, \&
  {Squires}}]{Borgani2001}
{Borgani}, S. {et~al.} 2001, \apj, 561, 13

\bibitem[{{Brown} {et~al.}(2005){Brown}, {Jannuzi}, {Dey}, \&
  {Tiede}}]{Brown2005}
{Brown}, M.~J.~I., {Jannuzi}, B.~T., {Dey}, A., \& {Tiede}, G.~P. 2005, \apj,
  621, 41

\bibitem[{{Bruzual} \& {Charlot}(2003)}]{bc2003}
{Bruzual}, G., \& {Charlot}, S. 2003, \mnras, 344, 1000

\bibitem[{{Chen} {et~al.}(2002){Chen}, {McCarthy}, {Marzke}, {Wilson},
  {Carlberg}, {Firth}, {Persson}, {Sabbey}, {Lewis}, {McMahon}, {Lahav},
  {Ellis}, {Martini}, {Abraham}, {Oemler}, {Murphy}, {Somerville}, {Beckett},
  \& {Mackay}}]{Chen2002}
{Chen}, H. {et~al.} 2002, \apj, 570, 54

\bibitem[{{Cimatti} {et~al.}(2002){Cimatti}, {Daddi}, {Mignoli}, {Pozzetti},
  {Renzini}, {Zamorani}, {Broadhurst}, {Fontana}, {Saracco}, {Poli},
  {Cristiani}, {D'Odorico}, {Giallongo}, {Gilmozzi}, \& {Menci}}]{Cimatti2002}
{Cimatti}, A. {et~al.} 2002, \aap, 381, L68

\bibitem[{{Coil} {et~al.}(2004){Coil}, {Davis}, {Madgwick}, {Newman},
  {Conselice}, {Cooper}, {Ellis}, {Faber}, {Finkbeiner}, {Guhathakurta},
  {Kaiser}, {Koo}, {Phillips}, {Steidel}, {Weiner}, {Willmer}, \&
  {Yan}}]{Coil2004}
{Coil}, A.~L. {et~al.} 2004, \apj, 609, 525

\bibitem[{{Colless} {et~al.}(2001){Colless}, {Dalton}, {Maddox}, {Sutherland},
  {Norberg}, {Cole}, {Bland-Hawthorn}, {Bridges}, {Cannon}, {Collins}, {Couch},
  {Cross}, {Deeley}, {De Propris}, {Driver}, {Efstathiou}, {Ellis}, {Frenk},
  {Glazebrook}, {Jackson}, {Lahav}, {Lewis}, {Lumsden}, {Madgwick}, {Peacock},
  {Peterson}, {Price}, {Seaborne}, \& {Taylor}}]{Colless2001}
{Colless}, M. {et~al.} 2001, \mnras, 328, 1039

\bibitem[{{Crist{\' o}bal-Hornillos} {et~al.}(2003){Crist{\' o}bal-Hornillos},
  {Balcells}, {Prieto}, {Guzm{\' a}n}, {Gallego}, {Cardiel}, {Serrano}, \&
  {Pell{\' o}}}]{Cristobal2003}
{Crist{\' o}bal-Hornillos}, D., {Balcells}, M., {Prieto}, M., {Guzm{\' a}n},
  R., {Gallego}, J., {Cardiel}, N., {Serrano}, {\' A}., \& {Pell{\' o}}, R.
  2003, \apj, 595, 71

\bibitem[{{Cutri} {et~al.}(2003){Cutri}, {Skrutskie}, {Van Dyk}, {Beichman},
  {Carpenter}, {Chester}, {Cambresey}, {Evans}, {Fowler}, {Gizis}, {Howard},
  {Huchra}, {Jarrett}, {Kopan}, {Kirkpatrick}, {Light}, {Marsh}, {McCallon},
  {Schneider}, {Stiening}, {Sykes}, {Weinberg}, {Wheaton}, {Wheelock}, \&
  {Zacharias}}]{Cutri2003}
{Cutri}, R.~M. {et~al.} 2003, Explanatory Supplement to the 2MASS All Sky Data
  Release (Pasadena: IPAC),
  http://www.ipac.caltech.edu/2mass/releases/allsky/doc/explsup.html

\bibitem[{{Daddi} {et~al.}(2001){Daddi}, {Broadhurst}, {Zamorani}, {Cimatti},
  {R{\" o}ttgering}, \& {Renzini}}]{Daddi2001}
{Daddi}, E., {Broadhurst}, T., {Zamorani}, G., {Cimatti}, A., {R{\"
  o}ttgering}, H., \& {Renzini}, A. 2001, \aap, 376, 825

\bibitem[{{Daddi} {et~al.}(2000){Daddi}, {Cimatti}, \& {Renzini}}]{Daddi2000}
{Daddi}, E., {Cimatti}, A., \& {Renzini}, A. 2000, \aap, 362, L45

\bibitem[{{de Vries} {et~al.}(2002){de Vries}, {Morganti}, {R{\" o}ttgering},
  {Vermeulen}, {van Breugel}, {Rengelink}, \& {Jarvis}}]{deVries2002}
{de Vries}, W.~H., {Morganti}, R., {R{\" o}ttgering}, H.~J.~A., {Vermeulen},
  R., {van Breugel}, W., {Rengelink}, R., \& {Jarvis}, M.~J. 2002, \aj, 123,
  1784

\bibitem[{{Dickinson} {et~al.}(2003){Dickinson}, {Papovich}, {Ferguson}, \&
  {Budav{\' a}ri}}]{Dickinson2003}
{Dickinson}, M., {Papovich}, C., {Ferguson}, H.~C., \& {Budav{\' a}ri}, T.
  2003, \apj, 587, 25

\bibitem[{{Djorgovski} {et~al.}(1995){Djorgovski}, {Soifer}, {Pahre}, {Larkin},
  {Smith}, {Neugebauer}, {Smail}, {Matthews}, {Hogg}, {Blandford}, {Cohen},
  {Harrison}, \& {Nelson}}]{Djorgovski1995}
{Djorgovski}, S. {et~al.} 1995, \apjl, 438, L13

\bibitem[{{Drory} {et~al.}(2001){Drory}, {Feulner}, {Bender}, {Botzler},
  {Hopp}, {Maraston}, {Mendes de Oliveira}, \& {Snigula}}]{Drory2001}
{Drory}, N., {Feulner}, G., {Bender}, R., {Botzler}, C.~S., {Hopp}, U.,
  {Maraston}, C., {Mendes de Oliveira}, C., \& {Snigula}, J. 2001, \mnras, 325,
  550

\bibitem[{{Eisenhardt} {et~al.}(2004){Eisenhardt}, {Stern}, {Brodwin}, {Fazio},
  {Rieke}, {Rieke}, {Werner}, {Wright}, {Allen}, {Arendt}, {Ashby}, {Barmby},
  {Forrest}, {Hora}, {Huang}, {Huchra}, {Pahre}, {Pipher}, {Reach}, {Smith},
  {Stauffer}, {Wang}, {Willner}, {Brown}, {Dey}, {Jannuzi}, \&
  {Tiede}}]{Eisenhardt2004}
{Eisenhardt}, P.~R. {et~al.} 2004, \apjs, 154, 48

\bibitem[{{Elston} {et~al.}(2003){Elston}, {Raines}, {Hanna}, {Hon}, {Julian},
  {Horrobin}, {Harmer}, \& {Epps}}]{Elston2003}
{Elston}, R., {Raines}, S.~N., {Hanna}, K.~T., {Hon}, D.~B., {Julian}, J.,
  {Horrobin}, M., {Harmer}, C.~F.~W., \& {Epps}, H.~W. 2003, in Instrument
  Design and Performance for Optical/Infrared Ground-based Telescopes. Edited
  by Iye, Masanori; Moorwood, Alan F. M. Proceedings of the SPIE, Volume 4841,
  1611--1624

\bibitem[{{Elston} {et~al.}(1988){Elston}, {Rieke}, \& {Rieke}}]{Elston1988}
{Elston}, R., {Rieke}, G.~H., \& {Rieke}, M.~J. 1988, \apjl, 331, L77

\bibitem[{{Elston} {et~al.}(1989){Elston}, {Rieke}, \& {Rieke}}]{Elston1989}
{Elston}, R., {Rieke}, M.~J., \& {Rieke}, G.~H. 1989, \apj, 341, 80

\bibitem[{{Ferguson} {et~al.}(2004){Ferguson}, {Dickinson}, {Giavalisco},
  {Kretchmer}, {Ravindranath}, {Idzi}, {Taylor}, {Conselice}, {Fall},
  {Gardner}, {Livio}, {Madau}, {Moustakas}, {Papovich}, {Somerville},
  {Spinrad}, \& {Stern}}]{Ferguson2004}
{Ferguson}, H.~C. {et~al.} 2004, \apjl, 600, L107

\bibitem[{{Finlator} {et~al.}(2000){Finlator}, {Ivezi{\' c}}, {Fan}, {Strauss},
  {Knapp}, {Lupton}, {Gunn}, {Rockosi}, {Anderson}, {Csabai}, {Hennessy},
  {Hindsley}, {McKay}, {Nichol}, {Schneider}, {Smith}, {York}, \& {the SDSS
  Collaboration}}]{Finlator2000}
{Finlator}, K. {et~al.} 2000, \aj, 120, 2615

\bibitem[{{Fontana} {et~al.}(2003){Fontana}, {Donnarumma}, {Vanzella},
  {Giallongo}, {Menci}, {Nonino}, {Saracco}, {Cristiani}, {D'Odorico}, \&
  {Poli}}]{Fontana2003}
{Fontana}, A. {et~al.} 2003, \apjl, 594, L9

\bibitem[{{Gardner}(1998)}]{Gardner1998}
{Gardner}, J.~P. 1998, \pasp, 110, 291

\bibitem[{{Gardner} {et~al.}(1993){Gardner}, {Cowie}, \&
  {Wainscoat}}]{Gardner1993}
{Gardner}, J.~P., {Cowie}, L.~L., \& {Wainscoat}, R.~J. 1993, \apjl, 415, L9

\bibitem[{{Gardner} {et~al.}(1996){Gardner}, {Sharples}, {Carrasco}, \&
  {Frenk}}]{Gardner1996}
{Gardner}, J.~P., {Sharples}, R.~M., {Carrasco}, B.~E., \& {Frenk}, C.~S. 1996,
  \mnras, 282, L1

\bibitem[{{Glazebrook} {et~al.}(1994){Glazebrook}, {Peacock}, {Collins}, \&
  {Miller}}]{Glazebrook1994}
{Glazebrook}, K., {Peacock}, J.~A., {Collins}, C.~A., \& {Miller}, L. 1994,
  \mnras, 266, 65

\bibitem[{{Haiman} {et~al.}(2001){Haiman}, {Mohr}, \& {Holder}}]{Haiman2001}
{Haiman}, Z., {Mohr}, J.~J., \& {Holder}, G.~P. 2001, \apj, 553, 545

\bibitem[{{Hoopes} {et~al.}(2003){Hoopes}, {Heckman}, {Jannuzi}, {Dey},
  {Brown}, {Ford}, \& {GALEX Science Team}}]{Hoopes2003}
{Hoopes}, C.~G., {Heckman}, T.~M., {Jannuzi}, B.~T., {Dey}, A., {Brown},
  M.~J.~I., {Ford}, A., \& {GALEX Science Team}. 2003, American Astronomical
  Society Meeting Abstracts, 203,

\bibitem[{{Houck} {et~al.}(2005){Houck}, {Soifer}, {Weedman}, {Higdon},
  {Higdon}, {Herter}, {Brown}, {Dey}, {Jannuzi}, {Le Floc'h}, {Rieke}, {Armus},
  {Charmandaris}, {Brandl}, \& {Teplitz}}]{Houck2005}
{Houck}, J.~R. {et~al.} 2005, \apjl, 622, L105

\bibitem[{{Huang} {et~al.}(1997){Huang}, {Cowie}, {Gardner}, {Hu}, {Songaila},
  \& {Wainscoat}}]{Huang1997}
{Huang}, J.-S., {Cowie}, L.~L., {Gardner}, J.~P., {Hu}, E.~M., {Songaila}, A.,
  \& {Wainscoat}, R.~J. 1997, \apj, 476, 12

\bibitem[{{Huang} {et~al.}(2001){Huang}, {Thompson}, {K{\" u}mmel},
  {Meisenheimer}, {Wolf}, {Beckwith}, {Fockenbrock}, {Fried}, {Hippelein}, {von
  Kuhlmann}, {Phleps}, {R{\" o}ser}, \& {Thommes}}]{Huang2001}
{Huang}, J.-S. {et~al.} 2001, \aap, 368, 787

\bibitem[{{Jannuzi} \& {Dey}(1999)}]{Jannuzi1999}
{Jannuzi}, B.~T., \& {Dey}, A. 1999, in ASP Conf. Ser. 191, Photometric
  Redshifts and the Detection of High Redshift Galaxies, ed. R. Weynmann, L.
  Storrie-Lombardi, M. Sawicki, \& R. Brunner (San Francisco; ASP), 111--117

\bibitem[{{Jenkins} {et~al.}(2001){Jenkins}, {Frenk}, {White}, {Colberg},
  {Cole}, {Evrard}, {Couchman}, \& {Yoshida}}]{Jenkins2001}
{Jenkins}, A., {Frenk}, C.~S., {White}, S.~D.~M., {Colberg}, J.~M., {Cole}, S.,
  {Evrard}, A.~E., {Couchman}, H.~M.~P., \& {Yoshida}, N. 2001, \mnras, 321,
  372

\bibitem[{{K{\" u}mmel} \& {Wagner}(2001)}]{Kummel2001}
{K{\" u}mmel}, M.~W., \& {Wagner}, S.~J. 2001, \aap, 370, 384

\bibitem[{{Kochanek} {et~al.}(2004){Kochanek}, {Eisenstein}, {Caldwell},
  {Cool}, {Green}, \& {AGES}}]{Kochanek2004}
{Kochanek}, C.~S., {Eisenstein}, D., {Caldwell}, N., {Cool}, R., {Green}, P.,
  \& {AGES}. 2004, American Astronomical Society Meeting Abstracts, 205,

\bibitem[{{Kochanek} {et~al.}(2001){Kochanek}, {Pahre}, {Falco}, {Huchra},
  {Mader}, {Jarrett}, {Chester}, {Cutri}, \& {Schneider}}]{Kochanek2001}
{Kochanek}, C.~S. {et~al.} 2001, \apj, 560, 566

\bibitem[{{Lonsdale} {et~al.}(2003){Lonsdale}, {Smith}, {Rowan-Robinson},
  {Surace}, {Shupe}, {Xu}, {Oliver}, {Padgett}, {Fang}, {Conrow},
  {Franceschini}, {Gautier}, {Griffin}, {Hacking}, {Masci}, {Morrison},
  {O'Linger}, {Owen}, {P{\' e}rez-Fournon}, {Pierre}, {Puetter}, {Stacey},
  {Castro}, {Del Carmen Polletta}, {Farrah}, {Jarrett}, {Frayer}, {Siana},
  {Babbedge}, {Dye}, {Fox}, {Gonzalez-Solares}, {Salaman}, {Berta}, {Condon},
  {Dole}, \& {Serjeant}}]{Lonsdale2003}
{Lonsdale}, C.~J. {et~al.} 2003, \pasp, 115, 897

\bibitem[{{Martini}(2001)}]{Martini2001}
{Martini}, P. 2001, \aj, 121, 598

\bibitem[{{McCarthy}(2004)}]{McCarthy2004}
{McCarthy}, P.~J. 2004, \araa, 42, 477

\bibitem[{{McLeod} {et~al.}(1995){McLeod}, {Bernstein}, {Rieke}, {Tollestrup},
  \& {Fazio}}]{McLeod1995}
{McLeod}, B.~A., {Bernstein}, G.~M., {Rieke}, M.~J., {Tollestrup}, E.~V., \&
  {Fazio}, G.~G. 1995, \apjs, 96, 117

\bibitem[{{Mignoli} {et~al.}(2005){Mignoli}, {Cimatti}, {Zamorani}, {Pozzetti},
  {Daddi}, {Renzini}, {Broadhurst}, {Cristiani}, {D'Odorico}, {Fontana},
  {Giallongo}, {Gilmozzi}, {Menci}, \& {Saracco}}]{Mignoli2005}
{Mignoli}, M. {et~al.} 2005, \aap, 437, 883

\bibitem[{{Minezaki} {et~al.}(1998){Minezaki}, {Kobayashi}, {Yoshii}, \&
  {Peterson}}]{Minezaki1998}
{Minezaki}, T., {Kobayashi}, Y., {Yoshii}, Y., \& {Peterson}, B.~A. 1998, \apj,
  494, 111

\bibitem[{{Moustakas} {et~al.}(1997){Moustakas}, {Davis}, {Graham}, {Silk},
  {Peterson}, \& {Yoshii}}]{Moustakas1997}
{Moustakas}, L.~A., {Davis}, M., {Graham}, J.~R., {Silk}, J., {Peterson},
  B.~A., \& {Yoshii}, Y. 1997, \apj, 475, 445

\bibitem[{{Murray} {et~al.}(2005){Murray}, {Kenter}, {Forman}, {Jones},
  {Green}, {Kochanek}, {Vikhlinin}, {Fabricant}, {Fazio}, {Brand}, {Brown},
  {Dey}, {Jannuzi}, {Najita}, {McNamara}, {Shields}, \& {Rieke}}]{Murray2005}
{Murray}, S.~S. {et~al.} 2005, astro-ph/0504084

\bibitem[{{Nagashima} {et~al.}(2002){Nagashima}, {Yoshii}, {Totani}, \&
  {Gouda}}]{nagashima2002}
{Nagashima}, M., {Yoshii}, Y., {Totani}, T., \& {Gouda}, N. 2002, \apj, 578,
  675

\bibitem[{{Pierre} {et~al.}(2004){Pierre}, {Valtchanov}, {Altieri}, {Andreon},
  {Bolzonella}, {Bremer}, {Disseau}, {Dos Santos}, {Gandhi}, {Jean}, {Pacaud},
  {Read}, {Refregier}, {Willis}, {Adami}, {Alloin}, {Birkinshaw}, {Chiappetti},
  {Cohen}, {Detal}, {Duc}, {Gosset}, {Hjorth}, {Jones}, {LeFevre}, {Lonsdale},
  {Maccagni}, {Mazure}, {McBreen}, {McCracken}, {Mellier}, {Ponman},
  {Quintana}, {Rottgering}, {Smette}, {Surdej}, {Starck}, {Vigroux}, \&
  {White}}]{Pierre2004}
{Pierre}, M. {et~al.} 2004, Journal of Cosmology and Astro-Particle Physics, 9,
  11

\bibitem[{{Rhoads} {et~al.}(2000){Rhoads}, {Malhotra}, {Dey}, {Stern},
  {Spinrad}, \& {Jannuzi}}]{Rhoads2000}
{Rhoads}, J.~E., {Malhotra}, S., {Dey}, A., {Stern}, D., {Spinrad}, H., \&
  {Jannuzi}, B.~T. 2000, \apjl, 545, L85

\bibitem[{{Roche} {et~al.}(2002){Roche}, {Almaini}, {Dunlop}, {Ivison}, \&
  {Willott}}]{Roche2002}
{Roche}, N.~D., {Almaini}, O., {Dunlop}, J., {Ivison}, R.~J., \& {Willott},
  C.~J. 2002, \mnras, 337, 1282

\bibitem[{{Rudnick} {et~al.}(2003){Rudnick}, {Rix}, {Franx}, {Labb{\' e}},
  {Blanton}, {Daddi}, {F{\" o}rster Schreiber}, {Moorwood}, {R{\" o}ttgering},
  {Trujillo}, {van de Wel}, {van der Werf}, {van Dokkum}, \& {van
  Starkenburg}}]{Rudnick2003}
{Rudnick}, G. {et~al.} 2003, \apj, 599, 847

\bibitem[{{Sheth} \& {Tormen}(1999)}]{Sheth1999}
{Sheth}, R.~K., \& {Tormen}, G. 1999, \mnras, 308, 119

\bibitem[{{Skrutskie} {et~al.}(1997){Skrutskie}, {Schneider}, {Stiening},
  {Strom}, {Weinberg}, {Beichman}, {Chester}, {Cutri}, {Lonsdale}, {Elias},
  {Elston}, {Capps}, {Carpenter}, {Huchra}, {Liebert}, {Monet}, {Price}, \&
  {Seitzer}}]{Skrutskie1997}
{Skrutskie}, M.~F. {et~al.} 1997, in ASSL Vol. 210, The Impact of Large Scale
  Near-IR Sky Surveys, eds. F. Garzon et al. (Dordrecht: Kluwer Academic
  Publishing), 25--32

\bibitem[{{Stanford} {et~al.}(2005){Stanford}, {Eisenhardt}, {Brodwin},
  {Gonzalez}, {Stern}, {Dey}, {Jannuzi}, {Brown}, \& {Elston}}]{Stanford2005}
{Stanford}, S.~A. {et~al.} 2005, submitted to \apj~Letters

\bibitem[{{Stanford} {et~al.}(1997){Stanford}, {Elston}, {Eisenhardt},
  {Spinrad}, {Stern}, \& {Dey}}]{Stanford1997}
{Stanford}, S.~A., {Elston}, R., {Eisenhardt}, P.~R., {Spinrad}, H., {Stern},
  D., \& {Dey}, A. 1997, \aj, 114, 2232

\bibitem[{{Steidel} {et~al.}(1998){Steidel}, {Adelberger}, {Dickinson},
  {Giavalisco}, {Pettini}, \& {Kellogg}}]{Steidel1998}
{Steidel}, C.~C., {Adelberger}, K.~L., {Dickinson}, M., {Giavalisco}, M.,
  {Pettini}, M., \& {Kellogg}, M. 1998, \apj, 492, 428

\bibitem[{{Steidel} {et~al.}(1999){Steidel}, {Adelberger}, {Giavalisco},
  {Dickinson}, \& {Pettini}}]{Steidel1999}
{Steidel}, C.~C., {Adelberger}, K.~L., {Giavalisco}, M., {Dickinson}, M., \&
  {Pettini}, M. 1999, \apj, 519, 1

\bibitem[{{Steidel} {et~al.}(2004){Steidel}, {Shapley}, {Pettini},
  {Adelberger}, {Erb}, {Reddy}, \& {Hunt}}]{Steidel2004}
{Steidel}, C.~C., {Shapley}, A.~E., {Pettini}, M., {Adelberger}, K.~L., {Erb},
  D.~K., {Reddy}, N.~A., \& {Hunt}, M.~P. 2004, \apj, 604, 534

\bibitem[{{Szokoly} {et~al.}(1998){Szokoly}, {Subbarao}, {Connolly}, \&
  {Mobasher}}]{Szokoly1998}
{Szokoly}, G.~P., {Subbarao}, M.~U., {Connolly}, A.~J., \& {Mobasher}, B. 1998,
  \apj, 492, 452

\bibitem[{{Totani} {et~al.}(2001){Totani}, {Yoshii}, {Maihara}, {Iwamuro}, \&
  {Motohara}}]{Totani2001}
{Totani}, T., {Yoshii}, Y., {Maihara}, T., {Iwamuro}, F., \& {Motohara}, K.
  2001, \apj, 559, 592

\bibitem[{{Wang} {et~al.}(2004){Wang}, {Khoury}, {Haiman}, \&
  {May}}]{WangS2004}
{Wang}, S., {Khoury}, J., {Haiman}, Z., \& {May}, M. 2004, \prd, 70, 123008

\bibitem[{{Wheelock} {et~al.}(1994){Wheelock}, {Gautier}, {Chillemi}, {Kester},
  {McCallon}, {Oken}, {White}, {Gregorich}, {Boulanger}, \&
  {Good}}]{Wheelock1994}
{Wheelock}, S.~L. {et~al.} 1994, NASA STI/Recon Technical Report N, 95, 22539

\bibitem[{{Yan} {et~al.}(2004){Yan}, {Thompson}, \& {Soifer}}]{Yan2004}
{Yan}, L., {Thompson}, D., \& {Soifer}, B.~T. 2004, \aj, 127, 1274

\bibitem[{{York} {et~al.}(2000){York}, {Adelman}, {Anderson}, {Anderson},
  {Annis}, {Bahcall}, {Bakken}, {Barkhouser}, {Bastian}, {Berman}, {Boroski},
  {Bracker}, {Briegel}, {Briggs}, {Brinkmann}, {Brunner}, {Burles}, {Carey},
  {Carr}, {Castander}, {Chen}, {Colestock}, {Connolly}, {Crocker}, {Csabai},
  {Czarapata}, {Davis}, {Doi}, {Dombeck}, {Eisenstein}, {Ellman}, {Elms},
  {Evans}, {Fan}, {Federwitz}, {Fiscelli}, {Friedman}, {Frieman}, {Fukugita},
  {Gillespie}, {Gunn}, {Gurbani}, {de Haas}, {Haldeman}, {Harris}, {Hayes},
  {Heckman}, {Hennessy}, {Hindsley}, {Holm}, {Holmgren}, {Huang}, {Hull},
  {Husby}, {Ichikawa}, {Ichikawa}, {Ivezi{\' c}}, {Kent}, {Kim}, {Kinney},
  {Klaene}, {Kleinman}, {Kleinman}, {Knapp}, {Korienek}, {Kron}, {Kunszt},
  {Lamb}, {Lee}, {Leger}, {Limmongkol}, {Lindenmeyer}, {Long}, {Loomis},
  {Loveday}, {Lucinio}, {Lupton}, {MacKinnon}, {Mannery}, {Mantsch}, {Margon},
  {McGehee}, {McKay}, {Meiksin}, {Merelli}, {Monet}, {Munn}, {Narayanan},
  {Nash}, {Neilsen}, {Neswold}, {Newberg}, {Nichol}, {Nicinski}, {Nonino},
  {Okada}, {Okamura}, {Ostriker}, {Owen}, {Pauls}, {Peoples}, {Peterson},
  {Petravick}, {Pier}, {Pope}, {Pordes}, {Prosapio}, {Rechenmacher}, {Quinn},
  {Richards}, {Richmond}, {Rivetta}, {Rockosi}, {Ruthmansdorfer}, {Sandford},
  {Schlegel}, {Schneider}, {Sekiguchi}, {Sergey}, {Shimasaku}, {Siegmund},
  {Smee}, {Smith}, {Snedden}, {Stone}, {Stoughton}, {Strauss}, {Stubbs},
  {SubbaRao}, {Szalay}, {Szapudi}, {Szokoly}, {Thakar}, {Tremonti}, {Tucker},
  {Uomoto}, {Vanden Berk}, {Vogeley}, {Waddell}, {Wang}, {Watanabe},
  {Weinberg}, {Yanny}, \& {Yasuda}}]{York2000}
{York}, D.~G. {et~al.} 2000, \aj, 120, 1579

\end{thebibliography}

\clearpage

\end{document}